\documentclass[aps,pra,notitlepage,reprint,superscriptaddress,nofootinbib]{revtex4-1}

\usepackage{amssymb}
\usepackage{amsfonts}
\usepackage{comment}
\usepackage{graphicx}
\usepackage{dcolumn}
\usepackage{bm}
\usepackage{amsmath}
\usepackage{upgreek}
\usepackage[colorlinks,linkcolor=red,citecolor=blue,urlcolor=red]{hyperref}
\usepackage[utf8]{inputenc}
\usepackage[T1]{fontenc}
\usepackage{mathptmx}
\usepackage{ulem}
\usepackage{physics}
\usepackage{xcolor}
\usepackage{lineno}
\newcommand\mbf[1]{\mathbf{#1}}
\newcommand{\argmax}{\mathop{\mathrm{argmax}}\limits}
\newcommand{\argmin}{\mathop{\mathrm{argmin}}\limits}

\AtBeginDocument{%
    \newwrite\bibnotes
    \def\bibnotesext{Notes.bib}
    \immediate\openout\bibnotes=\jobname\bibnotesext
    \immediate\write\bibnotes{@CONTROL{REVTEX41Control}}
    \immediate\write\bibnotes{@CONTROL{%
    apsrev41Control,author="08",editor="1",pages="1",title="0",year="1"}}
     \if@filesw
     \immediate\write\@auxout{\string\citation{apsrev41Control}}%
    \fi
}

\begin{document}

\title{Experimental Demonstration of a Quantum-Optimal Coronagraph Using Spatial Mode Sorters}

\author{Nico Deshler}
\altaffiliation{Authors contributed equally to this work.}
\affiliation{Wyant College of Optical Sciences, University of Arizona, Tucson, AZ 85721, USA}
\author{Itay Ozer}
\altaffiliation{Authors contributed equally to this work.}
\affiliation{Department of Electrical and Computer Engineering, University of Maryland, College Park, MD 20742, USA}
\author{Amit Ashok}
\affiliation{Wyant College of Optical Sciences, University of Arizona, Tucson, AZ 85721, USA}
\author{Saikat Guha}
\affiliation{Department of Electrical and Computer Engineering, University of Maryland, College Park, MD 20742, USA}

\date{\today}

\begin{abstract}
\noindent Deep sub-diffraction exoplanet discovery currently lies beyond the reach of state-of-the-art direct imaging coronagraphs, which typically have an inner working angle larger than the diffraction scale. We present an experimental demonstration of a direct imaging coronagraph design capable of achieving the quantum limits of exoplanet detection and localization below the Rayleigh diffraction limit. Our benchtop implementation performs a forward and inverse pass through a free-space programmable spatial mode sorter configured to isolate photons in a point spread function (PSF)-adapted mode basis. During the forward pass, the fundamental mode is rejected, effectively eliminating light from an on-axis point-like star. On the inverse pass, the remaining modes are coherently recombined to form an image of a faint companion. Our experimental system is shown localizing an artificial exoplanet at sub-diffraction distances from its host star under a 1000:1 star-planet contrast.
\end{abstract}

\maketitle

\section{Introduction}
\label{sec: Introduction}

The challenge of discovering habitable planets beyond our solar system has motivated astronomers to develop a diverse repertoire of exoplanet detection techniques. Broadly speaking, transit photometry, radial velocity, gravitational microlensing, and astrometry methods all monitor perturbations to the brightness, position, and spectrum of a prospective host star over time to infer the presence and dynamics of a faint orbiting companion \cite{Wright:2012_ExoplanetDetectionMethods}. While these methods have enjoyed great success in detecting exoplanets, contributing over 5,500 confirmed discoveries to date \cite{Akeson:2013,schneider:2011_ExoplanetEU}, they fundamentally rely on \textit{indirect} observations which provide limited information about more detailed planetary features. Remotely characterizing atmospheric composition, weather patterns, surface temperature, and surface gravity is crucial for understanding extrasolar chemical environments and identifying potential biosignatures \cite{Currie:DirectImagingSpectroscopy,Traub:2010_DirectImagingExo}. 

By comparison, direct imaging techniques aspire to spatially observe/resolve orbiting exoplanets, providing more comprehensive planetary data \cite{Seager:2010_ExoplanetAtmospheres,Biller:2018_ExoplanetAtmosphereImaging}. However, direct imaging faces two compounding obstacles: photon shot noise and diffraction. Exoplanets are extremely faint compared to their host stars, with relative brightness factors ranging from $10^{-5}$ for Hot Jupiters to $10^{-11}$ for Exo-Earths in the habitable zone. Moreover, the distance between an exoplanet and its host star often falls below the optical resolution capabilities of current space-based telescopes, residing in the so-called `sub-diffraction regime' \cite{Biller:2013}. Under these circumstances, light from the exoplanet overlaps with prominent diffraction features of the host star. This overlap, combined with the overwhelming shot noise generated by the bright star, effectively renders the exoplanet undetectable with conventional telescopes. Developments in coronagraphy techniques have succeeded in nulling an on-axis point-like star so that only light from the exoplanet reaches the detector \cite{Foo:2005,Mari:2012_SubRayleighVortex,Guyon:2010_PIAACMC,Aime:2002,Soummer:2003,Soummer:2005}. In this way, state-of-the-art coronagraphs suppress photon shot noise intrinsic to measuring classical states of light, thereby enhancing the signal-to-noise ratio (SNR) of exoplanet signatures \cite{Zurlo:2024_DirectImagingExo,Guyon:2006}.

\begin{figure}
    \centering
    \includegraphics[width=\linewidth]{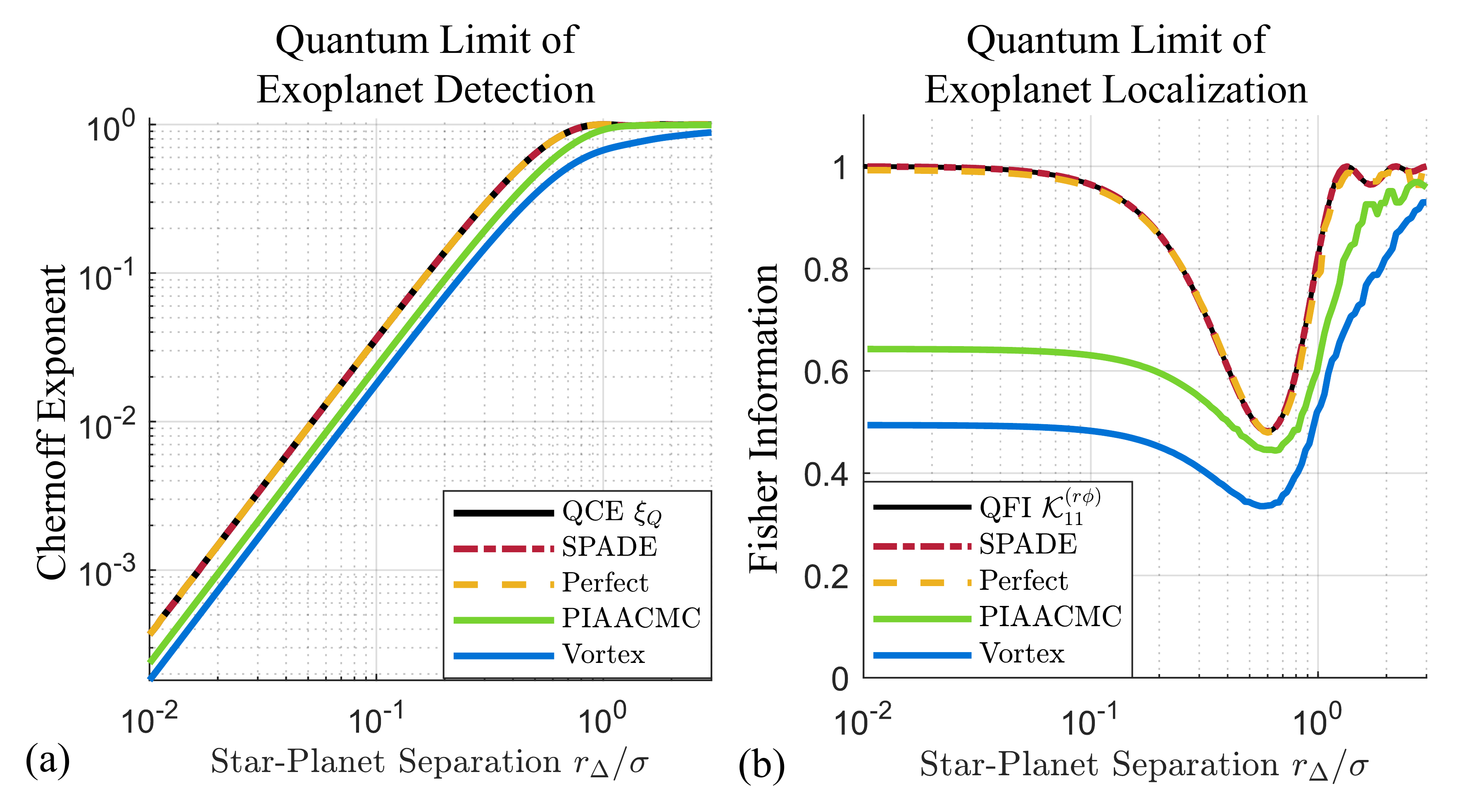}
    \caption{Comparisons between the Chernoff exponent \textbf{(a)} and the Fisher information per photon \textbf{(b)} as a function of the normalized star-planet separation $r_{\Delta}/\sigma$ for various coronagraph designs relative to the quantum limits (black) of exoplanet detection and localization at high contrasts. The Perfect Coronagraph (dashed yellow) is the focus of this work and is known to theoretically saturate both quantum limits in \cite{Deshler:2024}. State-of-the-art systems such as the phase-induced amplitude apodization complex mask coronagraph (PIAACMC) \cite{Guyon:2010_PIAACMC} and the vortex coronagraph \cite{Foo:2005} are notably sub-optimal in the sub-diffraction regime. Spatial mode demultiplexing (SPADE) \cite{Tsang:2016_QuantSuperresolution} is an alternative quantum-optimal measurement shown for reference, though it does not qualify as a direct imaging coronagraph.}
    \label{fig: Quantum Limits}
\end{figure}

Inspired by new insights in passive super-resolution imaging \cite{Tsang:2016_QuantSuperresolution,Tsang:2019_ResolvingStarlight,KitLee:2023_adaptiveSPADE,Prasad:3DSources}, we recently reported the quantum information limits for exoplanet detection and localization \cite{Deshler:2024}. Our findings revealed that these limits are achieved by a direct-imaging coronagraph that exclusively rejects the fundamental mode of the telescope. In contrast, current state-of-the-art coronagraphs discard information-bearing photons in higher-order modes \cite{Belikov:2021}, resulting in sub-optimal performance over the sub-diffraction regime, as illustrated in Figure \ref{fig: Quantum Limits}. Quantum-optimal coronagraphs, however, preserve information at sub-diffraction star-planet separations, where an abundance of exoplanets are expected to reside given  current statistical models \cite{Lagrange:2023_Radial,Fernandes:2019,Chen:2017,Ning:2018}.

Emerging spatial mode demultiplexing (SPADE) coronagraph platforms such as photonic lanterns \cite{Xin:2022,Xin:2023_PhotonicLanternCoronagraphExperiment} offer an alternative sensing paradigm that involves measuring optical power in different spatial modes. Although this approach provides a promising direction for exoplanet spectroscopy, it departs from the imaging paradigm, which we argue remains desirable in coronagraphy. In brief, images immediately provide the context and composition of a scene, thereby relaxing the amount of prior information needed to interpret and process measurements. Consider, for example, a prospective host star surrounded by multiple luminous sources (e.g. systems of exoplanets or exozodiacal dust). The distribution of these sources around the star is impossible to uniquely determine from modal power measurements alone, without additionally measuring the relative phase between modes or sorting multiple mode bases \cite{Kim:2024,Tsang:2019_IncoherentImaging_Part1,Tsang:2021_IncoherentImaging_Part2}. In imaging, such information is more readily apparent.

In this work, we propose a direct imaging coronagraph using a spatial mode sorter implemented with a multi-plane light converter (MPLC) \cite{Carpenter:2020,Fontaine:2019_LGModeSorter,Labroille:2014_ModeMultiplexerMPLC}. To the best of our knowledge, this is the first experimental verification of a coronagraph design that theoretically saturates the quantum limits for exoplanet discovery tasks. Applying a maximum likelihood estimator (MLE) to the images collected with our benchtop setup, we localize an artificial exoplanet at sub-diffraction separations from an artificial host star with a contrast ratio of 1000:1. Denoting the Rayleigh diffraction limit of our imaging system as $\sigma$,  our experiment achieves an absolute localization error below $0.03\sigma$ over the separation range $[0,\,0.6]\sigma$ with the MLE. 


\section{Methods}
\label{sec: Method}

\begin{figure}
    \centering
    \includegraphics[width=\linewidth]{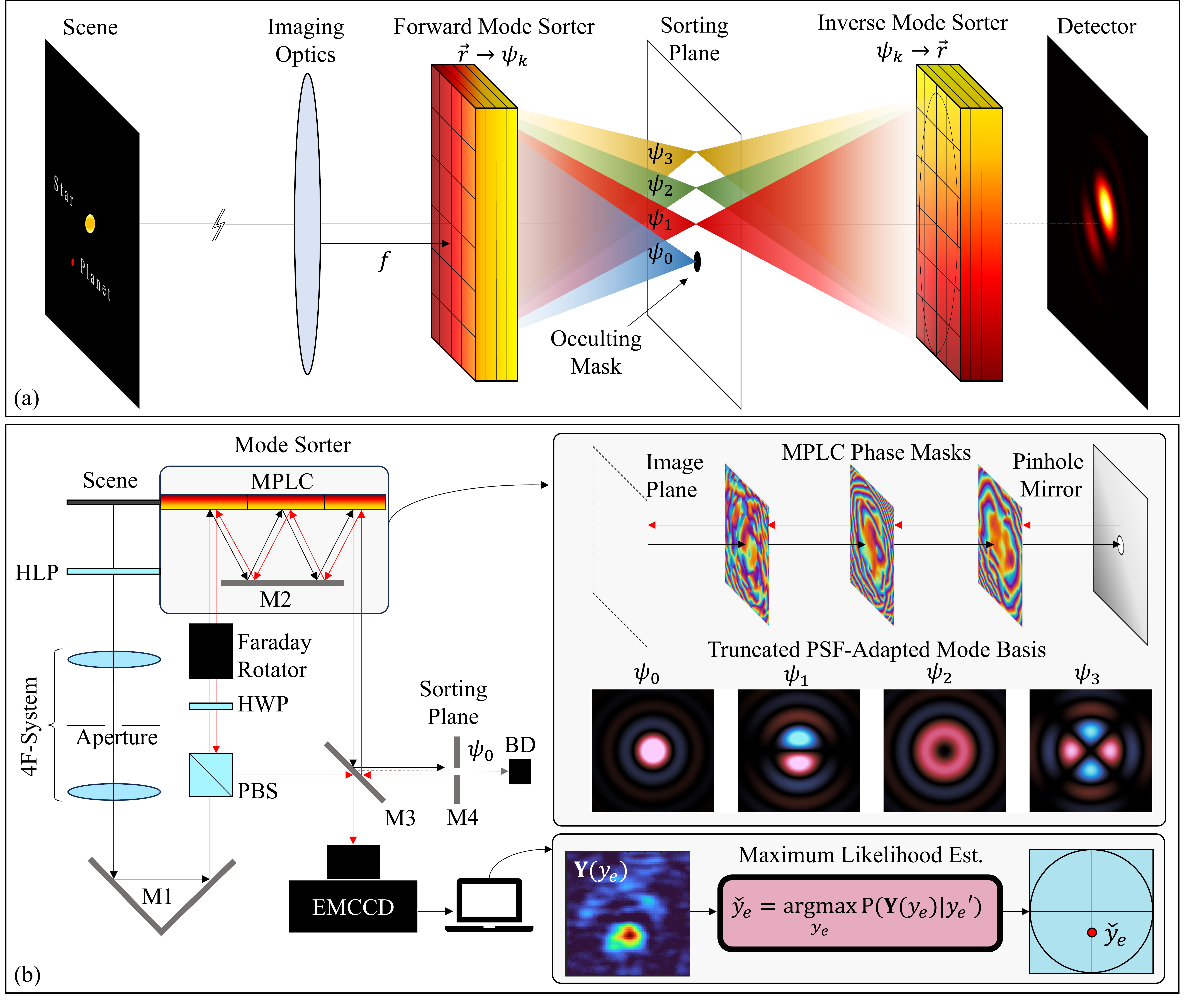}
    \caption{\textbf{(a)} Conceptual design of a quantum-optimal direct-imaging coronagraph based on spatial mode sorting. The forward mode sorter demultiplexes the field incident on the image plane into a PSF-adapted basis. Light in the fundamental mode is rejected by a mask at the sorting plane while light in the remaining modes propagates freely to an inverse mode sorter. The inverse mode sorter coherently recombines (multiplexes) the remaining modes to form an image at the detector. \textbf{(b)} A simplified schematic of the experimental implementation of a quantum-optimal coronagraph using an MPLC mode sorter and non-reciprocal polarization elements. The MPLC is comprised of a sequence of diffractive phase elements designed to (de)multiplex Fourier-Zernike modes $\{\bm{\psi}_0,\bm{\psi}_1,\bm{\psi}_2,\bm{\psi}_3\}$ where $\bm{\psi}_0$ is the PSF of the imaging system. A maximum likelihood estimator is applied on the captured image to localize the position of a sub-diffraction exoplanet.} 
    \label{fig: System Working Principle}
\end{figure}

\subsection{Experimental Design}
\label{sec: Experimental Design}


Figure \ref{fig: System Working Principle}(a) illustrates the working principles of a quantum-optimal coronagraph design based on two cascaded mode sorters. The first mode sorter decomposes the incident optical field into a PSF-adapted transverse spatial mode basis \cite{Rehacek:2018_MultiParameter_2Source} in order to isolate and eliminate photons residing in the fundamental mode. The second sorter inverts the mode decomposition, coherently recombining light in the residual modes to form an image of the exoplanet on a detector array. This scheme can be viewed as spatial mode filtering.

Our experimental setup shown in Figure \ref{fig: System Working Principle}(b) emulates the working principles of Figure \ref{fig: System Working Principle}(a) by double-passing the optical field through a single mode sorter implemented on a 3-plane MPLC detailed in \cite{Ozer:2022,Ozer:2024}. During the forward pass, the MPLC spatially demultiplexes the optical field in the Fourier-Zernike modes $\{\bm{\psi}_{nm}(\vec{r})\}$ (defined in Appendix \ref{apd: Fourier-Zernike}), which constitute a PSF-adapted basis for circular apertures, and focuses light in each mode to spatially resolved Gaussian spots on the sorting plane. The spot corresponding to the fundamental mode is directed to the opening of a pinhole mirror and absorbed at a beam dump. The remaining modes reflect off the pinhole mirror and are sent backwards through the mode sorter. The unitary nature of spatial mode sorting reverses the mode transformation during the backward pass. Non-reciprocal polarization elements split the optical path for the forward (pre-nulling) and backward (post-nulling) pass, sending the filtered field to a detector. Through this process, the field at the detector plane is identical to the field at the focal plane minus optical contribution from the fundamental mode. The $4f$ imaging system used in our setup is characterized by a circular aperture diameter $D = 400 \mu \text{m}$ and a focal length $f = 200 \text{mm}$ operating at wavelength $\lambda = 532 \text{nm}$, yielding a Rayleigh resolution of $\sigma = 1.22\lambda f/D = 324 \mu\text{m}$ on the object plane.

We focus on demonstrating exoplanet localization at sub-Rayleigh star-planet separations; the regime where quantum-optimal coronagraphs offer the greatest theoretical advantage over existing high-performance coronagraph designs. To sample this separation regime, we align the coronagraph to a bright on-axis point-source (artificial star) and vertically step the position of a second dim point-source (artificial exoplanet) $\vec{r}_e = (0,y_e)$ over the discrete domain $y_e \in \mathcal{Y} = [-h:\Delta:+h]$ with endpoints $h = 0.85 \sigma$ and sampling step size $\Delta = 0.0215 \sigma$.

\begin{figure}
    \centering
    \includegraphics[width=\linewidth]{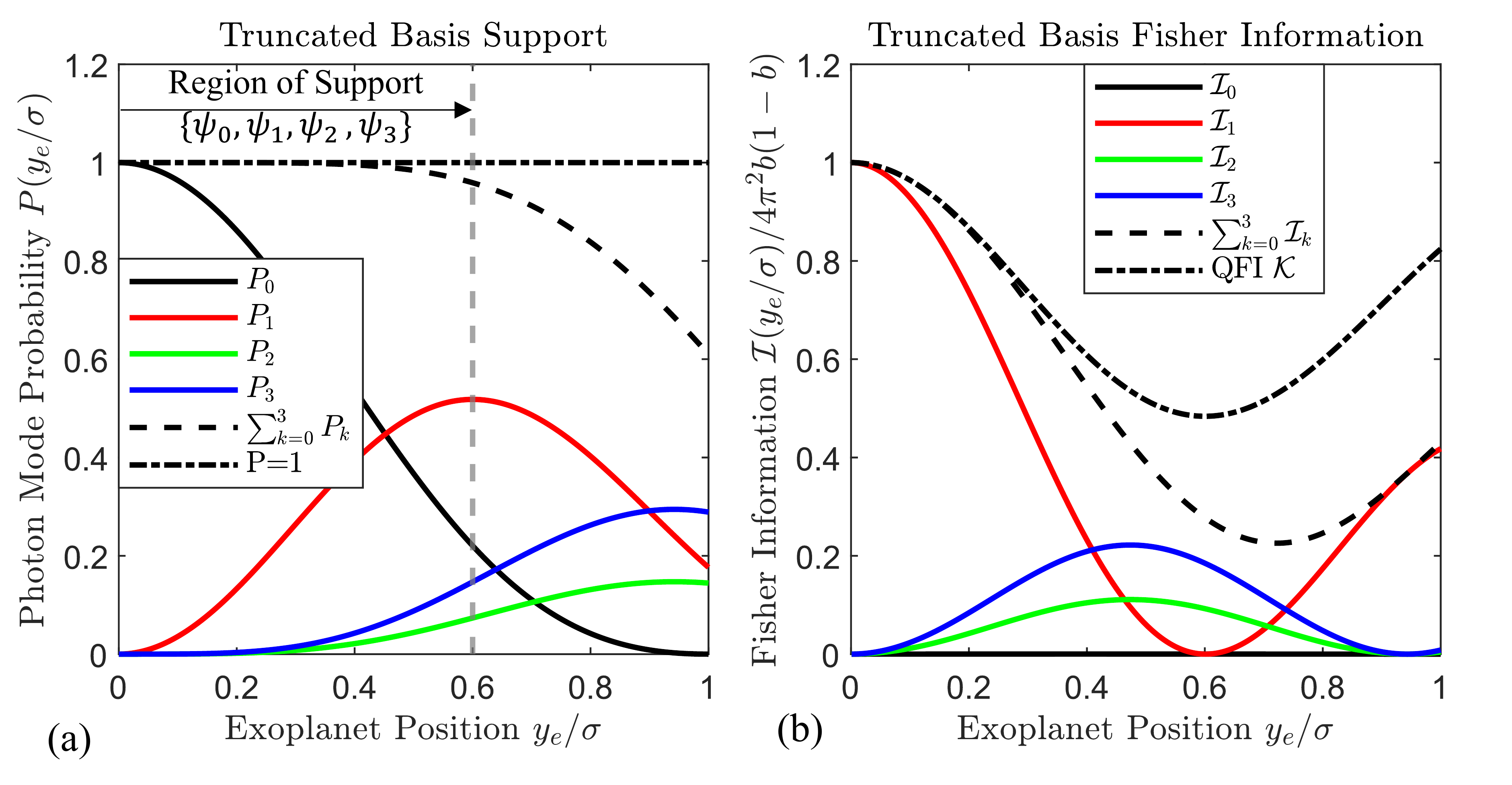}
    \caption{\textbf{(a)} Theoretical photon detection probabilities (coupling strengths) in each mode as a function of off-axis source location for the truncated set of Fourier-Zernike modes in the absence of cross-talk. The nominal 'region of support' is given by off-axis positions in the domain $[0, 0.6]\sigma$ within which $>95\%$ of the light energy couples to the truncated mode basis. \textbf{(b)} Distribution of the classical Fisher information in each mode as a function of off-axis position. The classical Fisher information of the truncated mode basis remains within $10\%$ of the quantum Fisher information limit over the domain $\sim [0, 0.4]\sigma$. Outside this domain, the information concentrates in higher-order modes that are ignored by our mode-sorter.}
    \label{fig: 4 Mode Zernike}
\end{figure}

Light from a sub-diffraction exoplanet couples predominantly to a small set of lower-order Fourier-Zernike modes. We therefore configured the MPLC to sort a truncated basis $\{ \bm{\psi}_0,\bm{\psi}_1,\bm{\psi}_2,\bm{\psi}_3\} = \{\bm{\psi}_{00},\bm{\psi}_{1-1},\bm{\psi}_{20},\bm{\psi}_{22}\}$ depicted in Figure \ref{fig: System Working Principle}(b) where $\bm{\psi}_0(\vec{r})$ is the fundamental mode of the imaging system. Collectively, these modes contain a majority of the energy in the field generated by a sub-Rayleigh companion traversing the $y$-axis as shown in Figure \ref{fig: 4 Mode Zernike}(a). We define the nominal region of support for this truncated basis to be $|y_e| \leq 0.6 \sigma$. For any exoplanet located in this region, >95\% of the optical energy couples to the truncated mode basis. 

The mode set used in our experimental setup was chosen specifically for 1D localization of a sub-diffraction exoplanet displaced along the y-axis. However, it is straightforward to configure the setup for 2D localization by augmenting/modifying the truncated basis to include modes which are sensitive to displacement along the x-axis. For example, we could have chosen to sort the modes $\{ \bm{\psi}_{00},\bm{\psi}_{1-1},\bm{\psi}_{11},\bm{\psi}_{20}\}$ which affords sub-diffraction sensitivity to horizontal displacement through the $\bm{\psi}_{11}$ mode. Expanding the basis to more than four modes was found to significantly degrade the cross-talk of the mode sorter due to the limited number of phase masks available on our programmable MPLC. In principle, introducing more masks would allow one to sort more modes and temper modal cross-talk, enabling access to larger star-planet separations and higher contrasts.

\subsection{Measurement Model}
\label{sec: Measurement Model}

\begin{figure}
    \centering
    \includegraphics[width=\linewidth]{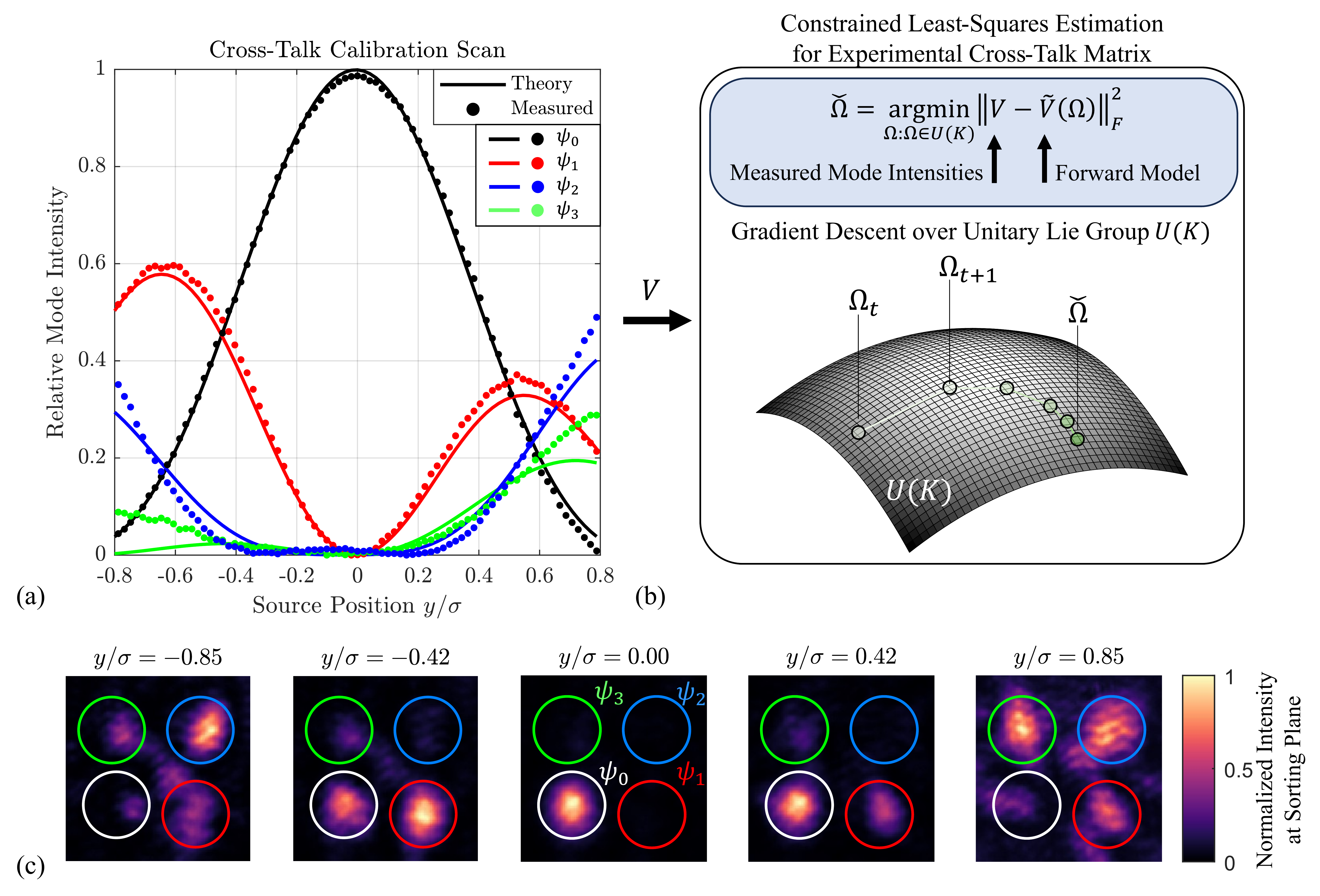}
    \caption{\textbf{(a)} Cross-talk calibration data (dotted) depicting the relative optical intensity measured in each target region of the sorting plane as a function of off-axis source position. Solid curves correspond to theoretically predicted mode intensities under a constrained least-squares fit of the unitary cross-talk matrix to the calibration data. \textbf{(b)} A visual representation of the least-squares fitting algorithm used to estimate the experimental cross-talk matrix under the constraint that $\Omega$ be a unitary matrix (further details in Appendix \ref{apd: Calibration}). \textbf{(c)} Example calibration images of the optical intensity distribution measured at the sorting plane for various off-axis source locations. The total intensity of each mode was taken to be the integrated intensity within their designated circular region on the detector.}
    \label{fig: Cross-Talk Calibration}
\end{figure}

\begin{figure}
    \centering
    \includegraphics[width=\linewidth]{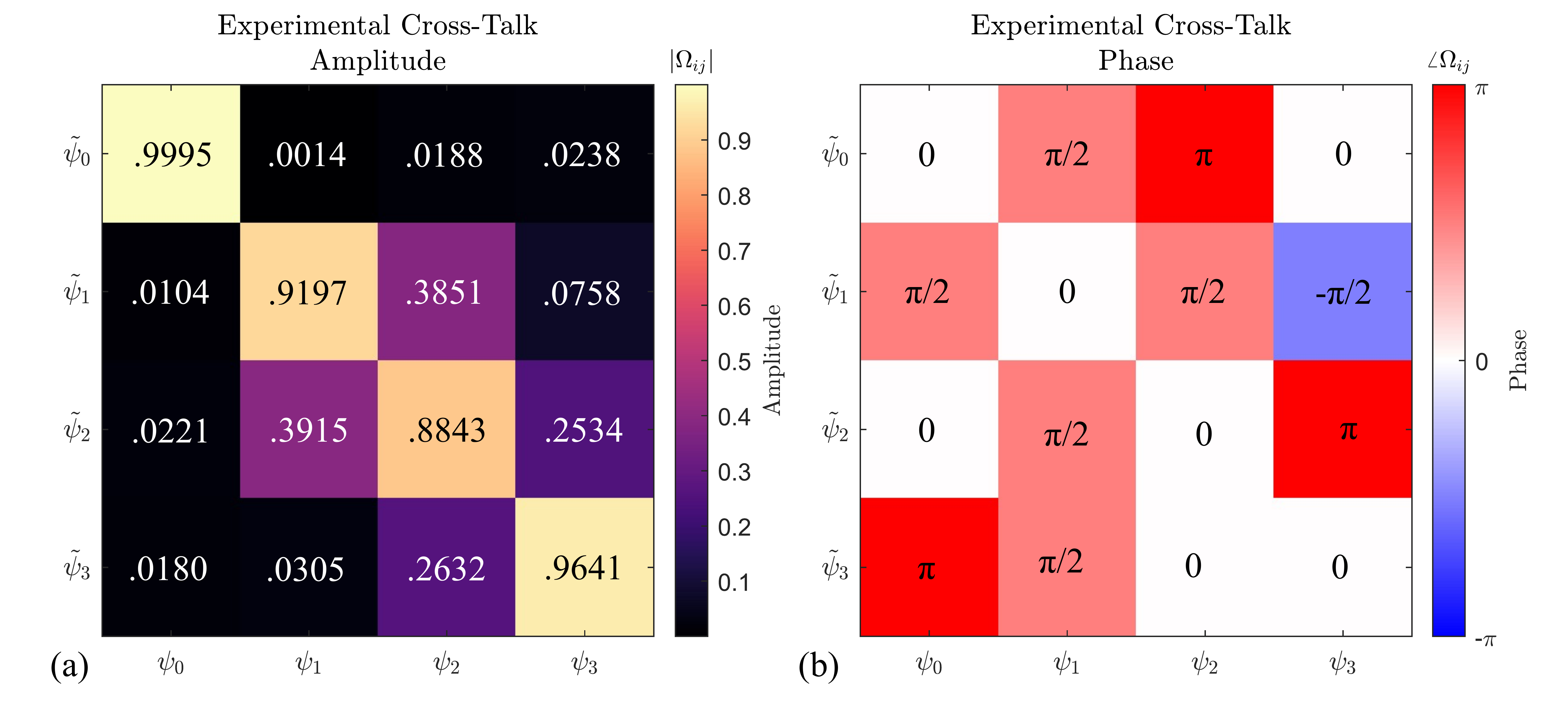}
    \caption{\textbf{(a)} Amplitude and \textbf{(b)} phase of the experimental cross-talk matrix recovered from the best-fit calibration method (depicted in Figure \ref{fig: Cross-Talk Calibration}) under the approximation that $\Omega$ be unitary. This approximation ignores scattering into modes beyond the span of the truncated mode set (see Appendix \ref{apd: Simulated Cross-Talk} for details).
    The first row of the amplitude matrix corresponds to leakage from the fundamental mode into higher order modes (imperfect star nulling), while the first column of the amplitude matrix corresponds to leakage from the higher order modes into the fundamental mode (lost exoplanet light). The phase components of the experimental cross-talk matrix are artifacts of the Fourier-Zernike mode definitions and the unitary approximation of $\Omega$. }
    \label{fig: Experimental Cross-Talk}
\end{figure}

\begin{figure}
\centering
    \includegraphics[width=\linewidth]{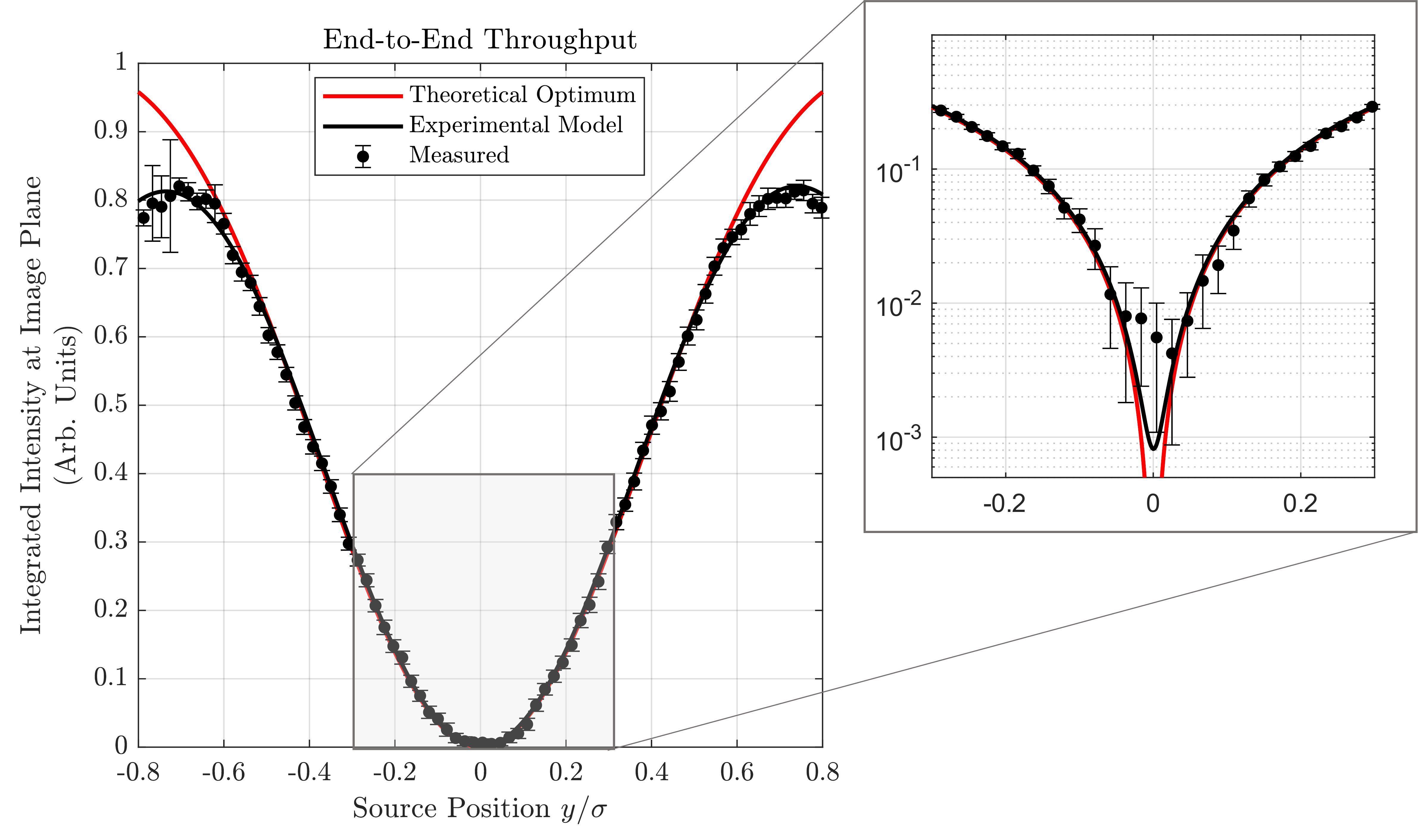}
    \caption{End-to-end throughput of the coronagraph as measured at the image plane (black scatter dots). The throughput predicted by the experimental model (solid black curve) makes use of the MPLC cross-talk matrix of Figure \ref{fig: Experimental Cross-Talk}. For reference, we also plot the ideal throughput (solid red curve) achieved by a perfect coronagraph that exclusively rejects the fundamental mode. The inset depicts the throughput on log-scale near the optical axis. When the source is on-axis, the measured throughput approximates the leakage from the fundamental mode predicted by the cross-talk matrix $\sim 10^{-3}$.}
    \label{fig: Throughput}
\end{figure}

\begin{figure*}
 \centering
 \includegraphics[width = \textwidth]{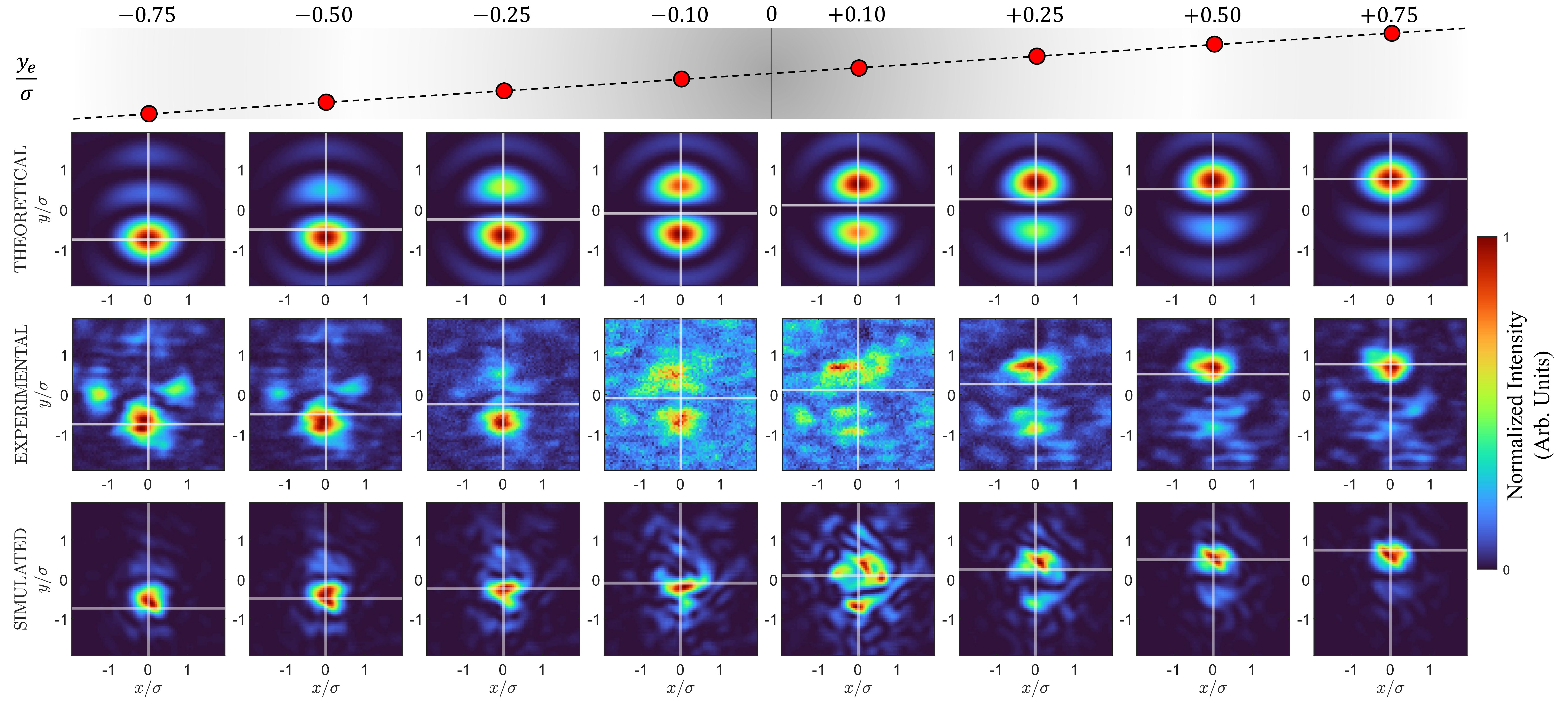}
 \caption{A comparison between theoretical, experimental, and simulated direct imaging measurements of the artificial exoplanet for various sub-diffraction star-planet separations along the y-axis under a 1000:1 star-planet contrast.  In each image, the ground truth location of the exoplanet coincides with the white cross-hairs while the star is located at the origin. All images are max-normalized to illustrate dominant spatial features. THEORETICAL: These images correspond to the perfect coronagraph operating in the absence of background or modal cross-talk. EXPERIMENTAL: These are images of the artificial exoplanet collected with our bench top coronagraph after subtraction of the background. SIMULATED: These images were generated by numerically propagating the optical field through a digital twin of the MPLC device used in our experimental setup. Noticeable distortions and asymmetries in the experimental and simulated images for equidistant off-axis $\pm y_{e}$ exoplanet locations arise from modal cross-talk matrices depicted in Figure \ref{fig: Experimental Cross-Talk}  and Figure \ref{fig: Simulated Cross-Talk} respectively.}
 \label{fig: Theory v Measurement}
\end{figure*}

We construct a statistical forward model for the images captured with our experimental coronagraph. To circumvent erroneous interference effects that would arise if both star and exoplanet were simultaneously illuminated by the coherent laser source in our setup, we synthesize a measurement of the star-planet scene by adding measurements of each body illuminated independently, $\mbf{Y} = \mbf{X}_s + \mbf{X}_e$. Here $\mbf{X}_s$ and $\mbf{X}_e$ are random vectors denoting image captured with our coronagraph when illuminating the artificial star and exoplanet respectively. Importantly, the statistics of the synthetic measurement $\mbf{Y}$ are identical to that expected for an incoherent star-planet pair. 

The $\mbf{X}_s$ and $\mbf{X}_e$ are themselves constructed from independent realizations of a single-shot measurement. Specifically, $\mbf{X}_s = \sum_{i = 1}^{N_s} \mbf{X}^{(i)}(\vec{0})$ and $\mbf{X}_e  = \mbf{X}(\vec{r}_e)$ where $\vec{r}_e$ is the position of the exoplanet and the choice of $N_s$ sets the star-planet contrast. Meanwhile, the single-shot measurement $\mbf{X}(\vec{r}_0) \in \mathbb{N}^M$ is a random vector denoting the number of photons measured at each pixel of a $M$-pixel detector array over fixed exposure time $T$ when imaging a single point-source located at position $\vec{r}_0$. The number of photons measured in each pixel are statistically independent Poisson random variables such that,
\begin{equation}
\mbf{X}(\vec{r}_0) \sim \text{Poiss}(\lambda_0 \mbf{q}(\vec{r}_0) +\lambda_B \mbf{p}_{B})
\label{eqn: Single-source Measurement Model}
\end{equation}
where $\lambda_0 \in \mathbb{R}$ is the photon flux entering the pupil from the point source, $\mbf{q}(\vec{r}_0) \in \mathbb{R}^{M}$ is the post-nulling photon arrival probability at each pixel after propagating the pupil field through the coronagraph, and $\mbf{p}_{B}\in \mathbb{R}^{M}$ is an experimentally-observed optical background distribution with flux rate $\lambda_B \in \mathbb{R}$ (Appendix \ref{apd: Measurement Model}). For simplicity, the flux rates $\lambda_0$ and $\lambda_B$ are expressed in photons per integration period $T$. The effect of the coronagraph on the incoming optical field manifests in the post-nulling photon arrival probability,
\begin{equation}
    \mbf{q}(\vec{r}_0) = |\Psi \Omega C \Omega^{\dagger} \Psi^{\dagger} \mbf{u}(\vec{r}_0)|^2,
    \label{eqn: single-source prob}
\end{equation}
where the input field $\mbf{u}(\vec{r}_0) =  \text{vec}[\bm{\psi}_0(\vec{r}-\vec{r}_0)] \in \mathbb{C}^{M}$ is a square-normalized discrete version of the shifted PSF induced by illumination from a point source located at $\vec{r}_0$ (Appendix \ref{apd: Function Vectorization}). The $|\cdot|^2$ operation is applied element-wise to convert the field at each pixel to intensity. We have introduced several system-dependent matrices that map between the image space of dimension $M$ and a truncated mode space of dimension $K$. In particular,  $\Psi \in \mathbb{C}^{M \times K}$ is a truncated change-of-basis matrix whose columns are the vectorized Fourier-Zernike modes expressed in the pixel basis (position representation), $\Omega \in \mathbb{C}^{K \times K}$ is the cross-talk matrix of the mode sorter, and $C \in \mathbb{R}^{K \times K}$ is a diagonal nulling matrix describing the rejection of the fundamental mode.

By the additive property of Poisson random variables, the complete statistical model for the synthetic measurement is,
\begin{equation}
\mbf{Y}(\vec{r}_e) \sim \text{Poiss}\bigg( \Lambda_0 \mbf{p}(\vec{r}_e) + \Lambda_B \mbf{p}_B\bigg),
\label{eqn: Measurement Model}
\end{equation}
where $\Lambda_0 = N\lambda_0$ and $\Lambda_B = N\lambda_B$ are the effective photon fluxes for the star-planet system and the background respectively, with $N=N_s + 1$ denoting the total number of measurement realizations. The photon distribution at the detector plane induced by the star-planet system is,
\begin{equation}
\mbf{p}(\vec{r}_e) = (1-b)\mbf{q}(\vec{0}) + b \mbf{q}(\vec{r}_e),
\label{eqn: star-planet photon distribution prob}
\end{equation}
where $b = 1/N$ is the relative brightness of the exoplanet. Thus the synthetic measurement emulates the statistics of imaging two unbalanced incoherent point sources simultaneously through the coronagraph. In our experimental setup: $K=4$, $M=77^2$, and $N_s = 1000$. Details regarding the empirical characterization of each parameter in the measurement model are provided in Appendix \ref{apd: Calibration}.

\subsection{Contrast Sensitivity Limit and Image Quality}
The cross-talk matrix $\Omega$ of the mode sorter determines the performance of our coronagraph in two critical ways: First, energy leakage into and out of the fundamental mode sets a heuristic limit on the maximum accessible star-planet contrast. Second, interference with residual light in the fundamental mode causes distortions to the image of the exoplanet. Figure \ref{fig: Experimental Cross-Talk} shows the amplitudes and phase of each entry in the cross-talk matrix $\Omega$.

The maximum star-planet contrast accessible with our benchtop coronagraph is set by the leakage of the fundamental mode into higher-order modes. This leakage causes light from the on-axis star to corrupt otherwise pristine signal from the exoplanet. The throughput of the coronagraph is given by $P(\vec{r}_0) =\mbf{1}^{\intercal}\mbf{q}(\vec{r}_0)$, which represents the probability that a photon emitted by a point source at position $\vec{r}_0$ arrives at the detector after propagating through the system. The leakage out of the fundamental mode is given by the throughput of the on-axis star $P_s = P(\vec{0})$. Equivalently, the throughput for an exoplanet located at $\vec{r}_e$ is given by $P_e = P(\vec{r}_e)$.

Under shot noise limited conditions, the SNR of the exoplanet signature is,
\begin{equation}
    SNR = \frac{\lambda_e P_e}{\sqrt{\lambda_e P_e + \lambda_s P_s}}
    \label{eqn: SNR}
\end{equation}
where $\lambda_e$ and $\lambda_s$ are the mean photon fluxes of the exoplanet and the star respectively over the measurement period. Note that the SNR is maximal when there is no leakage out of the fundamental mode $P_s=0$. Additionally, we note that the shot noise contributed by the residual star light and by the exoplanet light are of similar magnitudes when the leakage equals the relative brightness $P_s \approx \frac{\lambda_e}{\lambda_e + \lambda_s} = b$. For this reason, the leakage sets a heuristic limit to the accessible levels of star-planet contrast. Using the experimental cross-talk matrix depicted in Figure \ref{fig: Experimental Cross-Talk}, we find that the leakage for our benchtop system is $P_s = 9.1975\times 10^{-4}$ which sets a contrast limit of approximately $b \gtrsim 10^{-3}$. In Figure \ref{fig: Throughput}, we show the measured throughput for our coronagraph and compare it with the modeled throughput predicted by the experimental cross-talk matrix.

The entries of the cross-talk matrix corresponding to imperfect isolation of the fundamental mode (i.e. first row/column of $\Omega$) also introduce distortions to the final image of the exoplanet. To see this, first consider an ideal cross-talk matrix of the form
\begin{equation}
\Omega = 
\begin{bmatrix}
1 & \mbf{0}^{\intercal} \\
\mbf{0} &  \tilde{\Omega}
\end{bmatrix}
\label{eqn: Cross-Talk Ideal}
\end{equation}
where $\tilde{\Omega}\in\mathbb{C}^{(K-1) \times (K-1)}$ is a unitary sub-matrix describing the cross-talk between modes in the orthogonal complement space of the fundamental mode. Then, the details of $\tilde{\Omega}$ have no effect on the final image of the exoplanet since,
\begin{equation}
\Omega C \Omega^{\dagger} = C
\label{eqn: Nulling Matrix Invariance}
\end{equation}
such that Equation \ref{eqn: single-source prob} reduces to $
\mbf{q}(\vec{r}_0) = |\bm{\Psi} C \bm{\Psi}^{\dagger}\mbf{u}(\vec{r}_0)|^2
$. This invariance of the nulling matrix $C$ under transformation by $\Omega$ reflects the fact that the PSF-adapted basis is not unique.

Conversely, if the mode sorter does not perfectly isolate the fundamental mode (i.e. $\Omega$ does not assume the form of Equation \ref{eqn: Cross-Talk Ideal}), then interference between the residual light in the fundamental mode and orthogonal modes cause distortions to the image of the exoplanet. The precise nature of these distortions depends on the specific amplitudes and phase of each off-diagonal entry in the cross-talk matrix given the target basis that the mode sorter is configured to (de)multiplex. In practice, Equation \ref{eqn: Nulling Matrix Invariance} must be approached with progressively higher fidelity to gain access to more extreme star-planet contrasts and lower image distortion. In Appendix \ref{apd: Simulated Cross-Talk} we perform a simulated analysis of the cross-talk matrix for the MPLC deployed in our experimental setup.

\section{Results}
\label{sec: Results}
\subsection{Exoplanet Localization}
In Figure \ref{fig: Theory v Measurement}, we compare theoretical, experimental, and simulated images of an artificial star-planet system captured with our coronagraph. For separations $|y_e| > 0.1\sigma$ there are strong qualitative similarities between the theoretical and experimental image intensity profiles, indicating that the exoplanet signal dominates over the background noise. However, at separations below this threshold, where the vast majority of exoplanet photons are discarded with the fundamental mode, the inherent noise in our experimental system becomes more prominent than the exoplanet signal. Additionally, the asymmetry observed between images of the exoplanet positioned at equal distances above and below the optical axis ($\pm y_e$) emerges due to the cross-talk between modes $\bm{\psi}_{1}$ and $\bm{\psi}_{2}$.

For each exoplanet location $y_e \in \mathcal{Y}$, we compiled repeated synthetic measurements $\{\mbf{Y}^{(i)}(y_e)\}$ for $i = 1,\ldots,(\ell=100)$ with exposure time $T = 100 \text{ ms}$. To localize the exoplanet we employ a maximum likelihood estimator (MLE), 
\begin{equation}
\check{y}^{(i)}_e(y_e) = \argmax_{ y_e' \in \mathcal{Y}} \log P\big( \mbf{Y}^{(i)}(y_e) | y_e'\big)
\end{equation}
We note that the MLE of the exoplanet position implicitly assumes \textit{a priori} knowledge of the star-planet contrast $b$ via the distribution on $\mbf{Y}$. In the idealized case where the cross-talk matrix takes the form of Equation \ref{eqn: Cross-Talk Ideal} such that all light from the star is eliminated, knowledge of the contrast is not required because the measurement only contains contribution from the exoplanet $\mbf{Y} \propto \mbf{X}_e$. In practice, any true implementation of a mode sorter will likely suffer from some amount of cross-talk that prohibits complete extinction of the fundamental mode. Consequently, the star-planet contrast becomes a nuisance parameter for exoplanet localization and detection tasks. This reality applies to any coronagraphy system with leakage out of the fundamental mode, not specifically our design. A possible solution is to jointly estimate the exoplanet position and the contrast, though this extension lies beyond the scope of the current work.

\begin{figure}
    \centering
    \includegraphics[width=\linewidth]{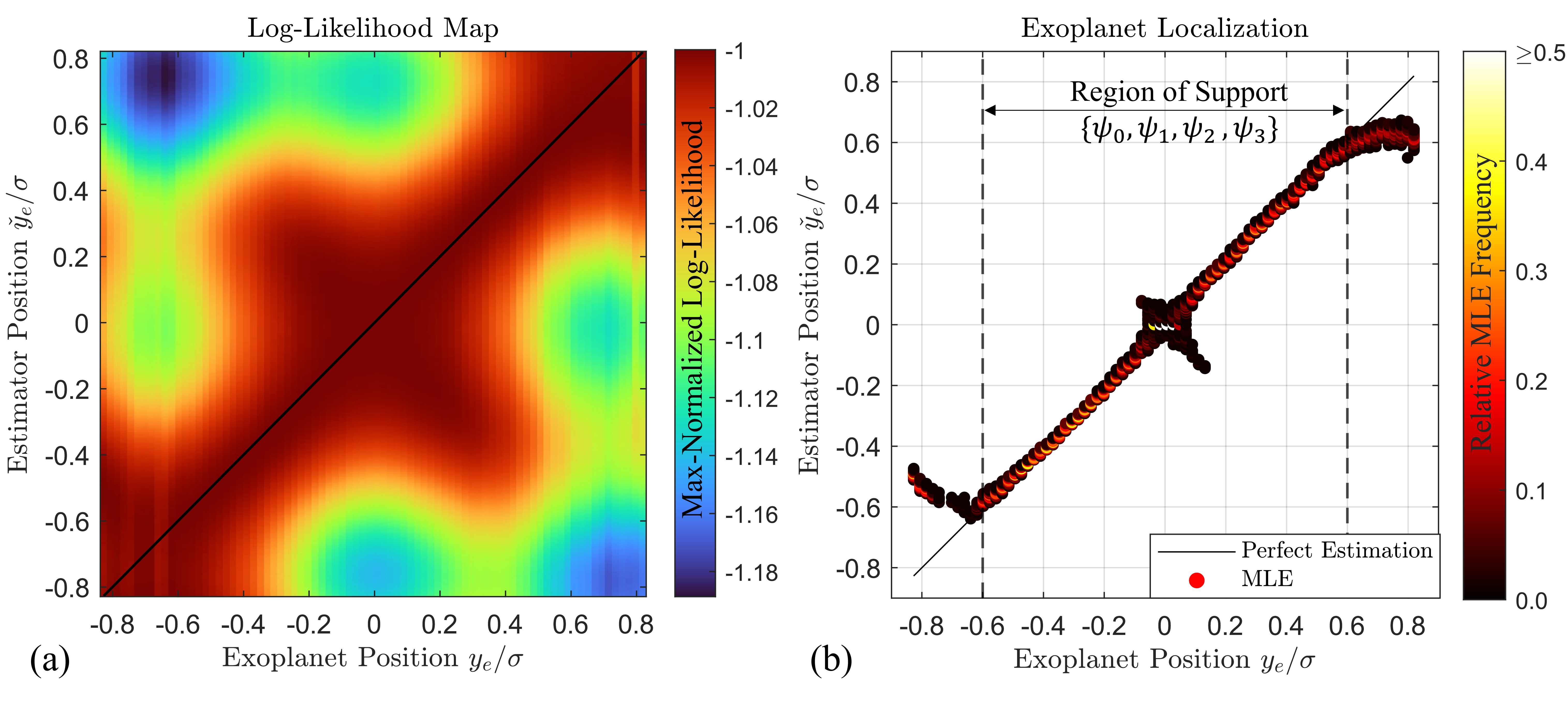}
    \caption{\textbf{(a)} Map of the log-likelihood $\mathcal{L}(\check{y}_{e};y_e)$ over sub-diffraction exoplanet locations. The peak likelihood ridge line (dark red regions) nearly coincides with perfect localization (black diagonal). However, certain regions near the optical axis and outside the support of the truncated mode basis exhibit weakly-peaked likelihoods indicating a greater variance in the MLE. \textbf{(b)} Scatter plot of the MLEs obtained from the complete measurement ensemble collected during the exoplanet position scan. The colorbar indicates the relative frequency of an estimate at each exoplanet position. Obvious outliers were removed for visual clarity.}
    \label{fig: Exoplanet Localizations}
\end{figure}

Figure \ref{fig: Exoplanet Localizations}(a) shows a map of the log-likelihood $\mathcal{L}(y_e;y_e') = \log P\big( \mbf{Y}(y_e) | y_e'\big)$ averaged over all experimental trials. The likelihood map exhibits a peak ridge line (maximum likelihood) that corresponds closely with the ground-truth position of the exoplanet. However, for exoplanets residing near the optical axis and outside the nominal region of support for the truncated mode basis, the likelihood exhibits weak curvature indicating greater estimator uncertainty. Figure \ref{fig: Exoplanet Localizations}(b) shows the MLEs obtained for all repeated measurements over the exoplanet translation scan, reinforcing salient features of the log-likelihood map. The MLE accurately localizes the exoplanet within the nominal region of support $|y_e| \leq .6\sigma$. Outside of this domain, the estimator experiences an increasing bias and uncertainty as photons are lost to modes beyond the support of the truncated mode set.

\subsection{Statistical Performance Analysis}

To further quantify the performance of our coronagraph we analyze the statistical error and imprecision across the MLE realizations $\{\check{y}_{e}^{(i)}(y_e) : y_{e} \in \mathcal{Y}\}$. Given a ground truth exoplanet position $y_{e}$, we denote the mean and variance of the MLE to be $\bar{y}_e(y_{e}) \equiv \mathbb{E}_{\mbf{Y}|y_e}[\check{y}_e]$ and $\sigma^2_{e}(y_{e}) = \mathbb{V}_{{\mbf{Y}|y_e}}[\check{y}_e]$ respectively. The unbiased empirical estimators of the MLE's mean and variance are given by the statistical averages,
\begin{subequations}
\begin{align}
\check{\bar{y}}_e(y_e) &\equiv \frac{1}{\ell}\sum_{i=1}^{\ell} \check{y}_{e}^{(i)}(y_e)    \\
\check{\sigma}^2_{e} (y_{e}) &\equiv \frac{1}{\ell-1} \sum_{i=1}^{\ell}(\check{y}_{e}^{(i)}(y_e) - \check{\bar{y}}_e (y_e) )^2.
\end{align}
\end{subequations}
The bias (expected error) of the MLE $\epsilon(y_e) = \mathbb{E}_{\mathbf{Y}|y_{e}}[\check{y}_{e}]- y_{e}$ also has an associated estimator given by,
\begin{equation}
\check{\epsilon}(y_{e}) = \check{\bar{y}}_e(y_e) - y_e
\end{equation}

\begin{figure}
    \centering
    \includegraphics[width=\linewidth]{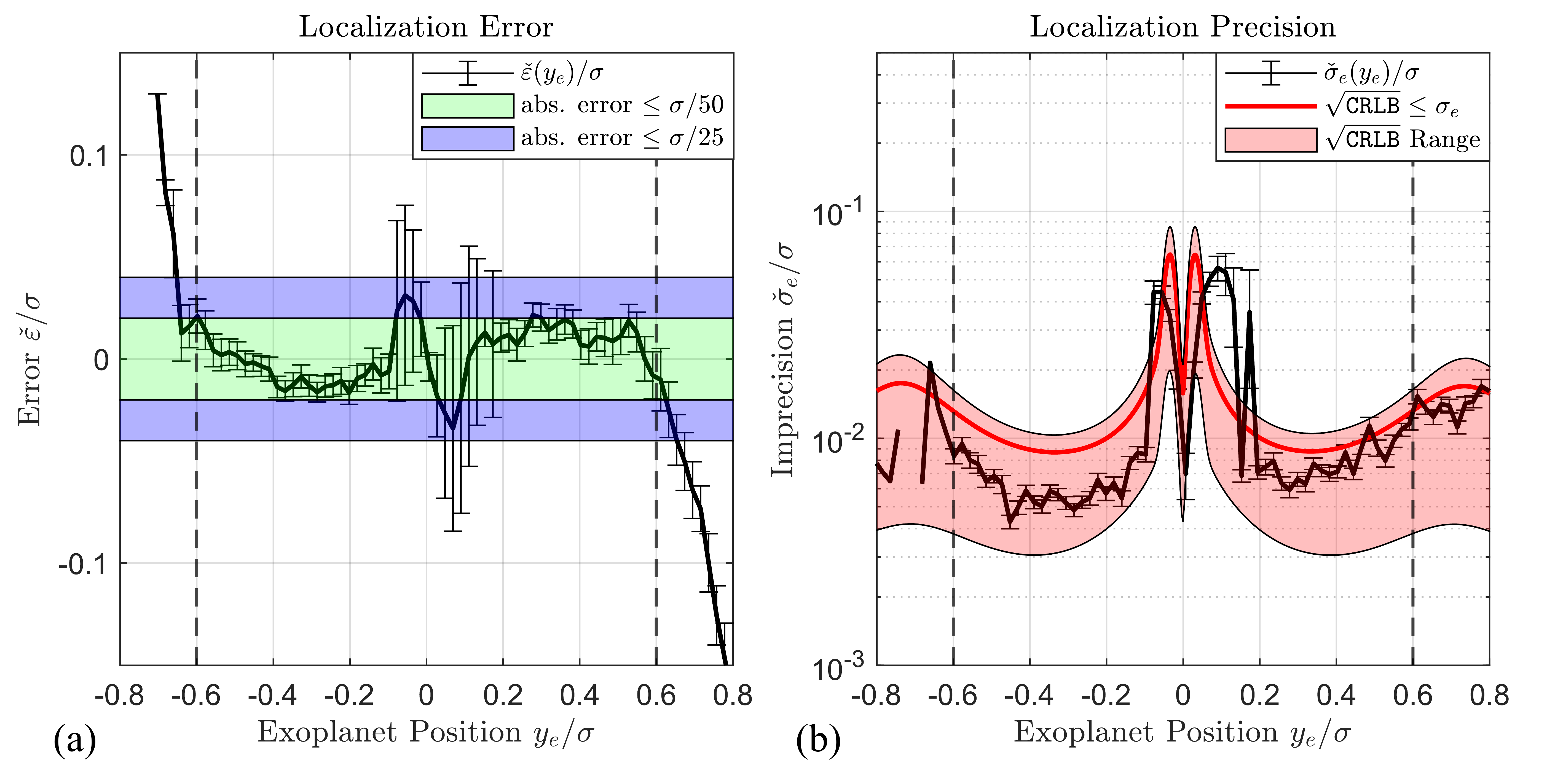}
    \caption{\textbf{(a)} Average error of the exoplanet MLE as a function of the ground truth position (black curve). Green and blue shaded regions demarcate where the absolute localization error is below $\sigma/50$ and $\sigma/25$ respectively. \textbf{(b)} Statistical standard deviation of the MLE as a function of exoplanet position as compared to the Cram\'er-Rao lower bound (CRLB) for the measurement model of Equation \ref{eqn: Measurement Model}. Uncertainty bars on the imprecision estimator were calculated using jackknife resampling \cite{Wu:1986_jackknife}. The red shaded region indicates viable bounds of the CRLB given uncertainty in the experimental model parameters. In particular we show the upper and lower bounds of the CRLB associated with the left and right sides of FWHM, respectively, for $\lambda_0$ in Figure \ref{fig: Lambda0 Estimates} (Appendix \ref{apd: Calibration}).}
    \label{fig: Error Analysis}
\end{figure}

Figure \ref{fig: Error Analysis}(a) shows the bias estimator (statistical error) of the MLE over the complete exoplanet position scan. In the subdomain $|y_e| \in [0.1,0.6]\sigma$, the absolute error is $|\check{\epsilon}(y_e)|<0.02\sigma$ with an imprecision $\check{\sigma}_{e}(y_e) \approx 0.01\sigma$, showcasing the localization accuracy achievable with our benchtop coronagraph at sub-Rayleigh star-planet separations. In the subdomain $|y_e|\in [0.01,0.1]\sigma$, where the experimental data is dominated by systematic error (primarily background and modal cross-talk), the absolute error is $|\check{\epsilon}(y_e)|<0.03\sigma$ with a relatively large imprecision $\check{\sigma}_{e}(y_e) \approx 0.1$ reflecting the fundamental uncertainty due to small signal in presence of significant measurement noise.

The variance of the MLE is subject to the Cram\'er-Rao lower bound (CRLB) for a biased estimator,
\begin{equation}
\sigma^2_e(y_e) \geq \frac{(1+\epsilon'(y_e))^2}{I(y_e)} = \text{CRLB}(y_e)
\label{eqn: CRLB}
\end{equation}
where $\epsilon'(y_e) = \frac{d\epsilon}{dy_e}$ is the derivative of the bias and $I(y_e)$ is the Fisher information of $y_e$ under the coronagraph measurement  model. This bound sets the minimum imprecision for the MLE. Figure \ref{fig: Error Analysis}(b) shows the empirical imprecision $\check{\sigma}_{e}(y_e)$ of the MLE as a function of the exoplanet position. We find that this imprecision curve corresponds well with the trends set by the CRLB of the MLE. As the exoplanet approaches the optical axis, a growing proportion of its photons couple to the fundamental mode and are discarded by the coronagraph. Thus, the majority of detected photons are supplied by the background, degrading the precision of the position estimate. Interestingly, when the exoplanet is nearly coincident with the optical axis, the MLE becomes a biased estimator (Appendix \ref{apd: CRLB Biased MLE}) giving rise to a prominent dip in the imprecision. Additionally, we also observe apparent secondary lobes in the imprecision beyond the region of support due to mode truncation error. These lobes correspond to a decrease in the Fisher information of the exoplanet position evident in Figure \ref{fig: 4 Mode Zernike}(b).

\section{Conclusion}
In summary, we present an experimental quantum-optimal direct imaging coronagraph design using spatial mode (de)multiplexing, which has inherent advantages for exoplanet exploration relative to indirect observation techniques. Furthermore, we demonstrate high-accuracy localization of an artificial exoplanet at sub-diffraction distances from its host star, which provides new capability to search and study exoplanets that may be inaccessible (i.e., in terms of inner working angle) to current state-of-the-art coronagraphs. 

The performance of our experimental system at deeply sub-diffraction separations is presently limited by a combination of modal cross-talk, detector dark noise, and ambient background light. These experimental non-idealities can be mitigated comprehensively with higher-fidelity detectors, mode-sorter, and further optical design optimizations, enabling operation at deeper sub-diffraction separations and higher contrast levels. For large star-planet separations our system performance is limited by the truncation of the mode basis, which may be improved by extending the number of phase planes comprising the MPLC. Furthermore, recent techniques for \textit{in situ} optimization of MPLC phase masks \cite{Rocha:2025} could overcome model mismatch between our experimental system and its digital counterpart, enabling faithful numerical performance analyses. Additionally, while we employ a synthetic measurement construction in this work, follow-on studies involving two simultaneously radiating sources would provide more robust validation of our coronagraph design. Overall, we believe our work provides a compelling basis for further exploration and development of mode-sorting solutions for high-contrast astronomical imaging tasks.

Looking forward, even-order coronagraphs may be implemented using spatial mode filtering techniques similar to those presented here to accommodate the finite size of host star. Previous works have shown that shot noise induced by an extended host star may be progressively tempered by judiciously filtering/attenuating higher-order optical modes \cite{Belikov:2021}. Furthermore, extending functionality to broadband sources will be critical for analyzing the spectrum of the exoplanet, which embeds key planetary science information. Broadband sources introduce mode mismatch such that the cross-talk matrix becomes wavelength-dependent. For sub-diffraction exoplanets, the constraints on acceptable levels of mode mismatch become more stringent \cite{Prasad:2022_Broadband,Grace_Thesis:2022} ostensibly limiting the efficacy of spatial mode sorting methods. However, recent numerical \cite{Zhang:2020} and experimental \cite{Mounaix:2020} work suggests that MPLCs may be used to simultaneously sort spectral and spatial modes, providing a common platform for exoplanet spectroscopy and/or nulling of a broadband star.

\section{Acknowledgments}
\label{sec: Acknowledgments}
\noindent ND acknowledges support from the National Science Foundation Graduate Research Fellowship under Grant No. DGE-2137419. IO and SG acknowledge that this research was supported by Raytheon and recognize the contributions of Jaime Bucay and Mark Meisner for their insights.
\section{Disclosures}
\label{sec: Disclosures}
\noindent The authors declare no conflicts of interest.
\section{Data Availability Statement}
\label{sec: Data Availability}
\noindent Our codebase for figure generation, coronagraph simulation, and measurement processing conducted in this work can be found in the project Github repository \cite{Deshler:2024_PerfectCoronagraph}. Raw measurement data files are hosted on Zenodo and are free to download at \cite{Deshler:2024_PerfectCoronagraphData}.

\newpage
\bibliography{references}

\appendix
\section{Fourier-Zernike Modes}
\label{apd: Fourier-Zernike}

For an imaging system with circular pupil of radius $R$, focal length $f$, and operating wavelength $\lambda$, let $(X_a,Y_a)$ and $(X_b,Y_b)$ denote the coordinate space of the pupil plane and focal plane respectively. We define the dimensionless pupil plane and focal plane coordinate vectors,
\begin{subequations}
\begin{align}
\vec{u} &\equiv \frac{1}{R}(X_a,Y_a) \\
\vec{r} &\equiv \frac{R}{\lambda f} (X_b,Y_b)
\end{align}
\end{subequations}

The Zernike modes $\tilde{\bm{\psi}}_{nm}(\vec{u})$ constitute a PSF-matched basis over a circular pupil. In polar coordinates, they are given by,

\begin{subequations}
    \begin{align}
    \tilde{\bm{\psi}}_{nm}(u,\theta) &\equiv  R_{nm}(u) \Theta_{m}(\theta) \text{circ}(u) \\
    R_{nm}(u) &\equiv \sum_{j=0}^{(n-|m|)/2} \frac{(-1)^j\sqrt{n+1}(n-j)! }{j![(n+m)/2 - j]![(n-m)/2 -j]!} u^{n-2j}\\
     \Theta_{m}(\theta) &\equiv
     \begin{cases}
     \sqrt{2} \cos(|m|\theta) & (m>0)\\
     1  &   (m = 0)\\
     \sqrt{2} \sin(|m|\theta) & (m<0)
     \end{cases}\\
     \text{circ}(u) &\equiv 
     \begin{cases}
     1, & u \leq 1\\
     0, & u > 1
     \end{cases}
    \end{align}
\label{eqn: Zernike Modes}
\end{subequations}
where the radial index range is $n = 0,1,2,\ldots,\infty$. The angular index range for a given radial index is $m \in S_n = \{-n,-n+2,\ldots,n-2,n\}$. These modes are defined to satisfy the orthonormality condition,
$$
\int \tilde{\bm{\psi}}^{*}_{nm}(u,\theta) \tilde{\bm{\psi}}_{n'm'} (u,\theta) udud\theta  = \delta_{nn'}\delta_{mm'}
$$
The Fourier transform of the Zernike modes over the pupil are found in \cite{Dai:06} to be,
\begin{equation}
\bm{\psi}_{nm}(r,\phi) = i^{n+2|m|}\sqrt{n+1} \frac{J_{n+1}(2\pi r)}{\sqrt{\pi} r}\Theta_{m}(\phi)
\label{eqn: Fourier-Zernike Modes}
\end{equation}
which we refer to as the 'Fourier-Zernike' modes throughout the main text. In this work, we sort the truncated mode basis $\{\bm{\psi}_0,\bm{\psi}_1,\bm{\psi}_{2},\bm{\psi}_3\}$ corresponding to modes $\{ \bm{\psi}_{00}, \bm{\psi}_{1,-1}, \bm{\psi}_{2,0}, \bm{\psi}_{2,2}\}$. The squared magnitude of these modes is shown in Figure \ref{fig: System Working Principle}c.

\section{Measurement Model}
\label{apd: Measurement Model}

\begin{figure}
\centering
\includegraphics[width=\linewidth]{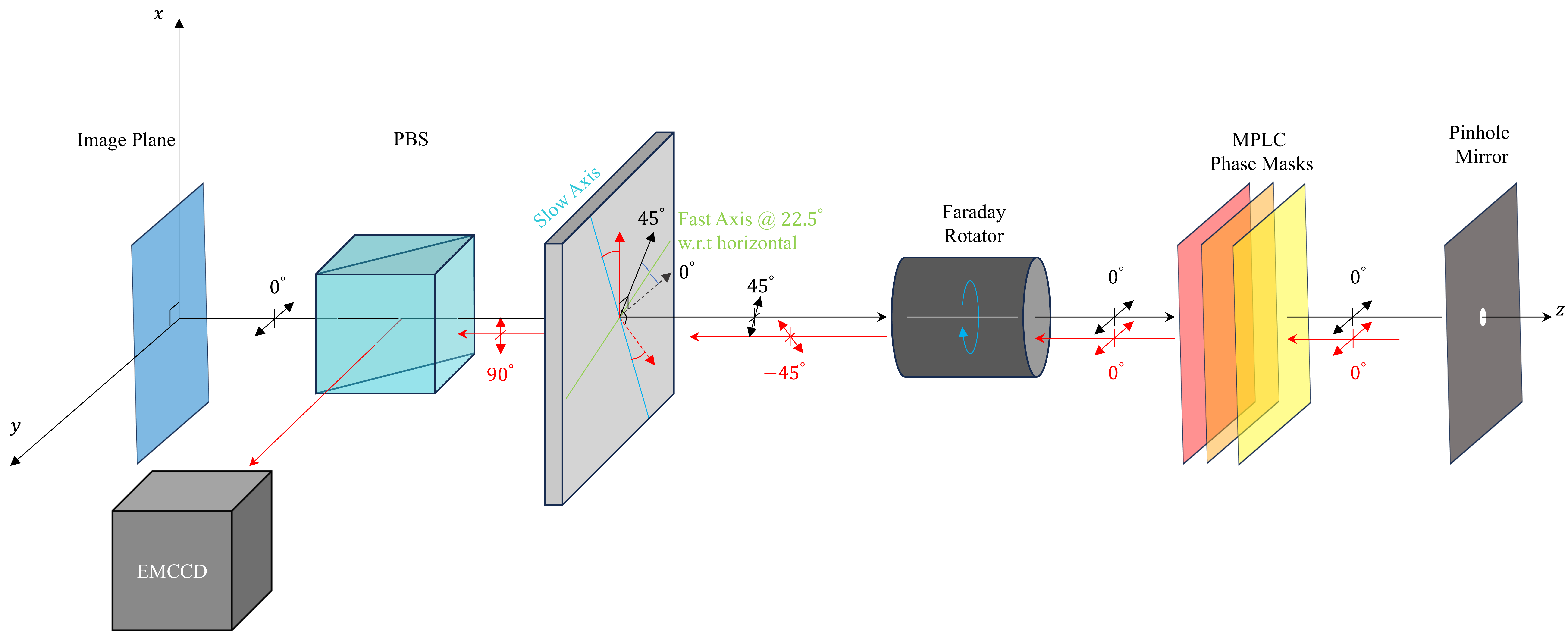}
\caption{Unfolded beam path illustrating polarization control for path splitting of forward (black arrows) and backward (red arrows) propagating beams. Through this manipulation we implement the design of Figure \ref{fig: System Working Principle} using a single mode sorter.}
\label{fig: Polarization Control}
\end{figure}

Here we expand on the single-shot measurement model of Equation \ref{eqn: Measurement Model}. The post-nulling optical intensity distribution on the detector is given in Equation \ref{eqn: single-source prob} to be, 
$$
\mbf{q}(\vec{r}_0) = |\Psi \Omega C \Omega^{\dagger} \Psi^{\dagger} \mbf{u}(\vec{r}_0)|^2
$$
which provides a matrix description of forward propagating the optical field through the MPLC, nulling the fundamental mode, back-propagating the field to an image plane, and measuring the intensity on a detector array. Here, $\mbf{u}(\vec{r}_0) = \text{vec}[\bm{\psi}(\vec{r}-\vec{r}_0)]$ is the vectorized form of the shifted PSF satisfying $\mbf{u}(\vec{r}_0)^{\dagger}\mbf{u}(\vec{r}_0) = 1$. $\Psi$ is the truncated change-of-basis matrix for which the columns are the vectorized Fourier-Zernike modes $\psi_{k} = \text{vec}[\bm{\psi}_k(\vec{r})]$
\begin{equation}
\Psi = 
\begin{bmatrix}
|& & | \\
\psi_0& \cdots & \psi_{K-1}\\
|& & |
\end{bmatrix}
\end{equation}
satisfying the orthonormality condition $\Psi^{\dagger}\Psi = I_{K}$ where $I_{K}$ is the identity matrix on $\mathbb{C}^{K}$. The coronagraph matrix is,
\begin{equation}
C = \pm
\begin{bmatrix}
0 & \mbf{0}^{\intercal} \\
\mbf{0} &  I_{K-1}\\
\end{bmatrix}
\label{eqn: Nulling Matrix}
\end{equation}
which serves to null the port assigned to the fundamental mode. The sign of coronagraph matrix depends on whether the system is transmissive $(+)$ at the sorting plane as in Figure \ref{fig: System Working Principle}(a) or reflective $(-)$ at the sorting plane as in our experimental configuration depicted in Figure \ref{fig: System Working Principle}(b). While we treat the optical field as a scalar quantity (as the field is always is in a definite linear polarization state throughout our experiment), details on the experimental polarization manipulation for path splitting are illustrated in Figure \ref{fig: Polarization Control}.

\section{Function Vectorization}
\label{apd: Function Vectorization}

Let $\Delta x, \Delta y$ be the horizontal and vertical pixel pitch dimensions for specific camera, and let $\{ (x_i,y_i)\}_{i=1}^{M}$ be the set of points corresponding to the center of each pixel. The $\text{vec}(\cdot)$ operation takes a well-behaved function $f(x,y)$ over the plane and converts it to a discrete column vector $\mbf{f}$ whose entries are given by,
\begin{equation}
    \text{f}_{i} = \int_{x_i -\Delta x/2}^{x_i +\Delta x/2}\int_{y_i -\Delta y/2}^{y_i +\Delta y/2}f(x,y) dx dy
\end{equation}

\section{Calibration}
\label{apd: Calibration}

This section addresses our procedures for characterizing all system-level parameters $\{\Omega,\lambda_D,\lambda_0,\mbf{s}\}$ taken to be known quantities in the measurement model.

\subsection{Cross-Talk Matrix $\Omega$}
Figure \ref{fig: Cross-Talk Calibration} shows calibration measurements made at the sorting plane of the MPLC for characterizing modal cross-talk. For each  position $y_i \in \mathcal{Y}$ of the point source, where $i=1,\ldots,(P=|\mathcal{Y}|)$, we measure the intensity in each mode channel $\mbf{v}_i \in \mathbb{R}^{K}$. We estimate the cross-talk matrix from these calibration measurements by optimizing a least-squares objective function $\mathcal{F}(\Omega)$ under the constraint that $\Omega$ be unitary,
\begin{equation}
    \check{\Omega} = \argmin_{\Omega \,\,: \,\,\Omega \in U(K)} \underbrace{||V - \tilde{V}(\Omega) ||_{F}^{2}}_{\mathcal{F}(\Omega)}
\end{equation}
where $||\cdot||_{F}$ is the Frobenius-norm and $\odot$ is the Hadamard element-wise product. In formulating this optimization problem, we define the following terms:

\begin{itemize}
    \item $V \in \mathbb{R}^{K \times P}$: Measurements of intensity in each mode channel. The $i^{th}$ column of $V$ is the measurement vector $\mathbf{v}_i$ collected for the $i^{th}$ position of the point source. 
    \item $\tilde{V}(\Omega) = (\Omega^{\dagger} Z)\odot(\Omega^{\dagger} Z)^{*}  \in  \mathbb{R}^{K \times P}:$ Forward model of the expected intensities measured in each mode channel under the unitary cross-talk matrix $\Omega$.
    \item $Z \in \mathbb{R}^{K \times P}$: Mode expansion coefficients for each off-axis source location in the calibration scan. The $i^{th}$ column of $Z$ is the vector $\mbf{z}_{i} = \Psi^\dagger\mbf{u}(y_i)\in\mathbb{C}^K$ for $y_i\in\mathcal{Y}$.
    \item $\Gamma_{\Omega} = \bigg(\frac{\partial \mathcal{F}}{\partial \Omega^{*}}\bigg)(\Omega) \equiv \frac{1}{2}\bigg[\frac{\partial \mathcal{F}}{\partial \Omega_{R}} + i \frac{\partial \mathcal{F}}{\partial \Omega_{I}}  \bigg]$: Gradient of the objective function with respect to the cross-talk matrix in the Euclidean space $\mathbb{R}^{2K\times 2K}$ where $\Omega_{R} = \Re{\Omega}$ and $\Omega_{I} = \Im{\Omega}$. The explicit Euclidean gradient under the squared error loss function is,
    \begin{equation}
         \bigg(\frac{\partial\mathcal{F}}{\partial \Omega^{*}} \bigg) (\Omega) = -2 Z\, [(V-\tilde{V}(\Omega))\odot(\Omega^{\dagger} Z)]^{\dagger} 
    \end{equation}
    \item $G(\Omega) = \Gamma_{\Omega} \Omega^{\dagger} - \Omega \Gamma^{\dagger}_{\Omega}$: Riemannian gradient of the objective function over the Lie group $U(K)$.
\end{itemize}

We solve this optimization using the constrained descent algorithm detailed in \cite{Abrudan:2008,Abrudan:2005_UnitaryMatrixOptimizatioN}. The algorithm is briefly summarized as follows. Instantiate $\Omega_0 =I$ with learning rate $\mu>0$. Then, at iteration $t$, we update the cross-talk matrix as,
\begin{equation}
    \Omega_{t+1} = \texttt{expm}(-\mu G_t) \Omega_t 
\end{equation}
where $G_t = G(\Omega_t)$ and $\texttt{expm}(\cdot)$ is a numerical matrix exponentiation function. The iterations are performed until convergence. In Figure \ref{fig: Cross-Talk Calibration}(a), we show the measured mode intensity data alongside the theoretical mode intensities after applying the best-fit cross-talk matrix shown in \ref{fig: Experimental Cross-Talk}.

\subsection{Dark Noise Rate $\lambda_D$ and Structured Background $\mbf{s}$}

\begin{figure}
    \centering
    \includegraphics[width=\linewidth]{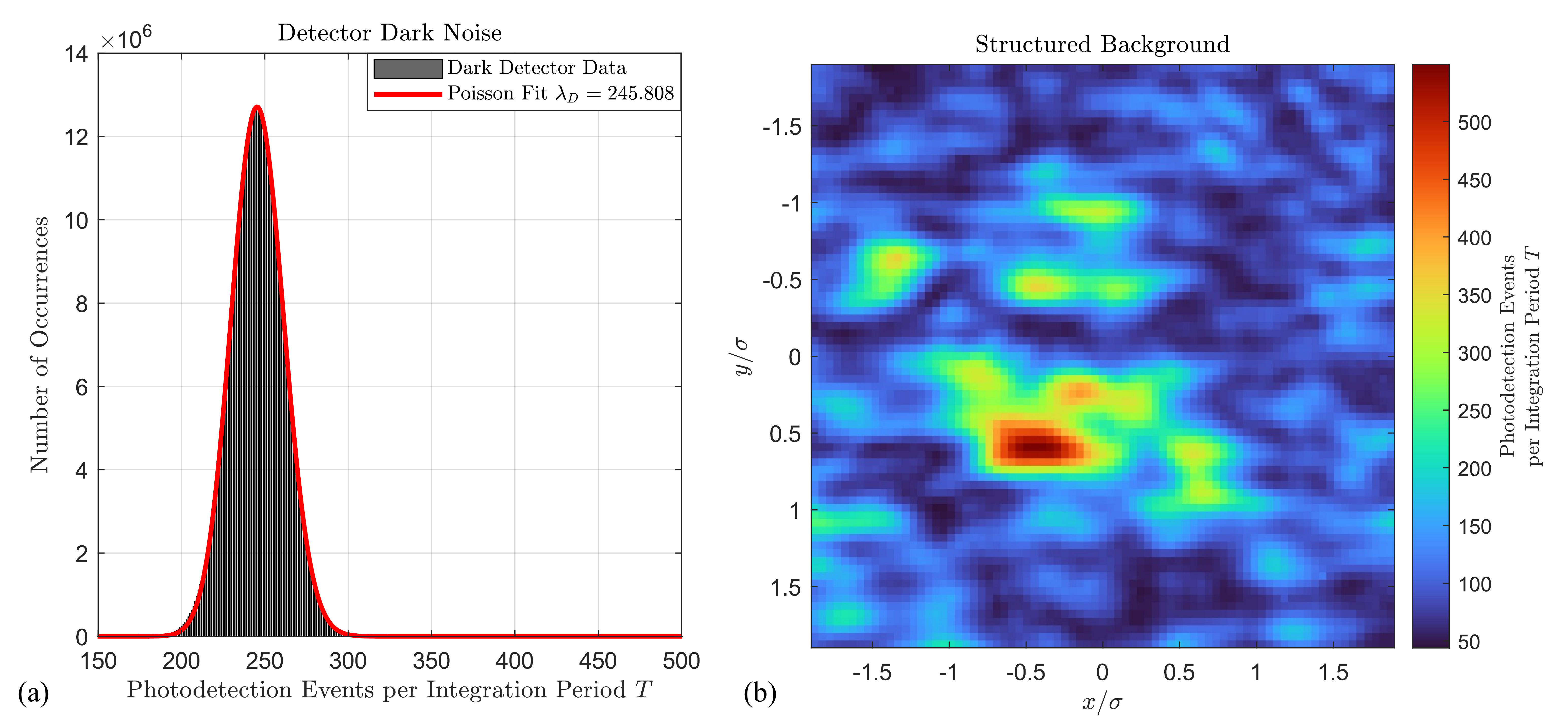}
    \caption{\textbf{(a)} Poisson fit to detector dark-noise histogram. The dark rate of the detector was found to be $\lambda_D = 246$ [photons $\cdot T^{-1}$] per pixel. \textbf{(b)} Structured background profile $\mbf{s}$ induced by stray light and undesired reflections in the experimental setup.}
    \label{fig: Noise Terms Calibration}
\end{figure}

The background contribution term employed in our single-shot measurement model arises from a combination of undesired stray light reflections and the dark noise rate of the pixels in the camera. We thus decompose the background into two terms as,
\begin{equation}
    \lambda_B \mbf{p}_B = \mbf{s} +\lambda_D \mbf{1}
    \label{eqn: Background Model}
\end{equation}
where $\mbf{s} \in \mathbb{R}^{M}$ is a structured background rate observed across our experimental measurements, and $\lambda_D \in \mathbb{R}$ is the spatially-uniform dark click rate of each pixel operating a room temperature. Furthermore, $\mbf{1}\in \mathbb{R}^{M}$ is a vector of ones, and we have impose the constraint $\mbf{1}^{\intercal}\mbf{p}_{B} = 1$ such that $\mbf{p}_{B}$ is a discrete probability distribution. Figure \ref{fig: Noise Terms Calibration} shows calibration data for estimating the dark noise rate $\lambda_D$ and structured background rate $\mbf{s}$ of Equation \ref{eqn: Background Model}. An estimate of the dark noise rate $\check{\lambda}_D = 246$ [photons/$T$] was found by fitting a Poisson distribution to a histogram of photon counts recorded by our camera operating in a dark environment at room temperature. The structured background rate was estimated by averaging $n=10^3$ images of the an on-axis source as viewed through the coronagraph,
\begin{equation}
    \check{\mbf{s}} = \frac{1}{n}\sum_{i=1}^{n} [\mbf{X}^{(i)}(\vec{0}) - \lambda_D\mbf{1}] 
\end{equation}
where we have assumed that the photons arriving at the detector due to imperfect nulling of the fundamental mode (i.e. modal cross-talk) are negligible compared to the dark noise rate $\lambda_0 \mbf{q}(\vec{0}) << \lambda_D$.

\subsection{Signal Photon Rate $\lambda_0$}

The mean photon rate entering the pupil from the star-planet scene is approximated using the Best Linear Unbiased Estimator (BLUE) \cite{Aitken:1936}. Let the random variable $Y^{(i)}_j(\vec{r}_e)$ represent the photons collected in the $j^{th}$ pixel of the $i^{th}$ measurement for a given exoplanet location $\vec{r}_e$. In accordance with equation \ref{eqn: Measurement Model}, each pixel follows distribution

$$
Y^{(i)}_j(\vec{r}_e) \sim \text{Poiss}\bigg( \Lambda_0 p_{j}(\vec{r}_e) + \Lambda_B p_{B_{j}} \bigg) 
$$
which we approximate as a non-random term linear in $\Lambda_0$ with zero-mean additive Gaussian noise,

\begin{align}
Y^{(i)}_j(\vec{r}_e) &= \Lambda_0 p_{j}(\vec{r}_e) + \Lambda_B p_{B_{j}} + \epsilon_{j} \\
\epsilon_{j}(\Lambda_0,\vec{r}_e) &\sim \mathcal{N}(0,\Lambda_0 p_{j}(\vec{r}_e) + \Lambda_B p_{B_j})
\end{align}

Defining the residuals, 

\begin{equation}
r_{j}^{(i)}(\Lambda_0, \vec{r}_e) = Y_{j}^{(i)}(\vec{r}_e) - \Lambda_0 p_j(\vec{r}_e) - \Lambda_B p_{B_{j}}
\end{equation}
the BLUE is found by optimizing the sum of squared residuals weighted by the variance of the photon arrivals in each pixel

\begin{equation}
\check{\Lambda}_0 = \argmax_{\Lambda_0} \sum_{y_e \in\mathcal{Y}}\sum_{i=1}^{\ell}\sum_{j=1}^{M} \frac{\big( r^{(i)}_{j}(\Lambda_0,y_e) \big)^2}{\mathbb{V}[\epsilon_j(\Lambda_0,y_e)]}.
\end{equation}

We numerically solve this optimization problem over the entire collected data set and estimate the mean photon rate entering the pupil from a single point source to be $\check{\lambda}_0 = \check{\Lambda}_0/N =  3.17\times10^6$ [photons/T].

\begin{figure}
    \centering
    \includegraphics[width=\linewidth]{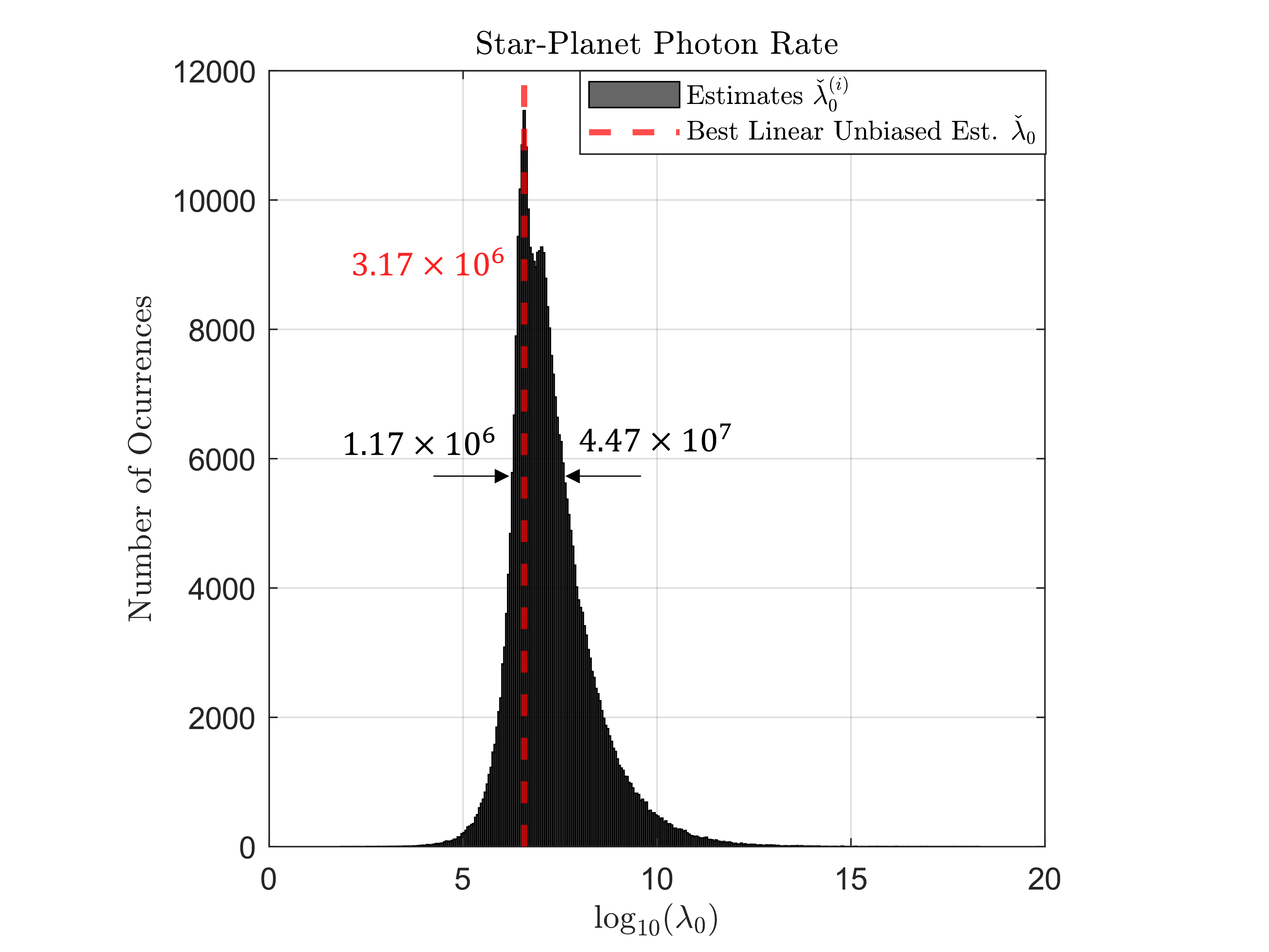}
    \caption{Histogram over estimates of the scene photon emission rate $\lambda_0$ from which we compute the best linear unbiased estimator (dashed red) $\check{\lambda}_0 = 3.17\times10^{6}$ [Photons $\cdot T^{-1}$]. The edges of the FWHM are used in Figure \ref{fig: Error Analysis} to determine the range of Cram\'er-Rao Bounds associated with our experiment.}
    \label{fig: Lambda0 Estimates}
\end{figure}

\subsection{Magnification and Alignment Correction}
\begin{figure}
    \centering
    \includegraphics[width=\linewidth]{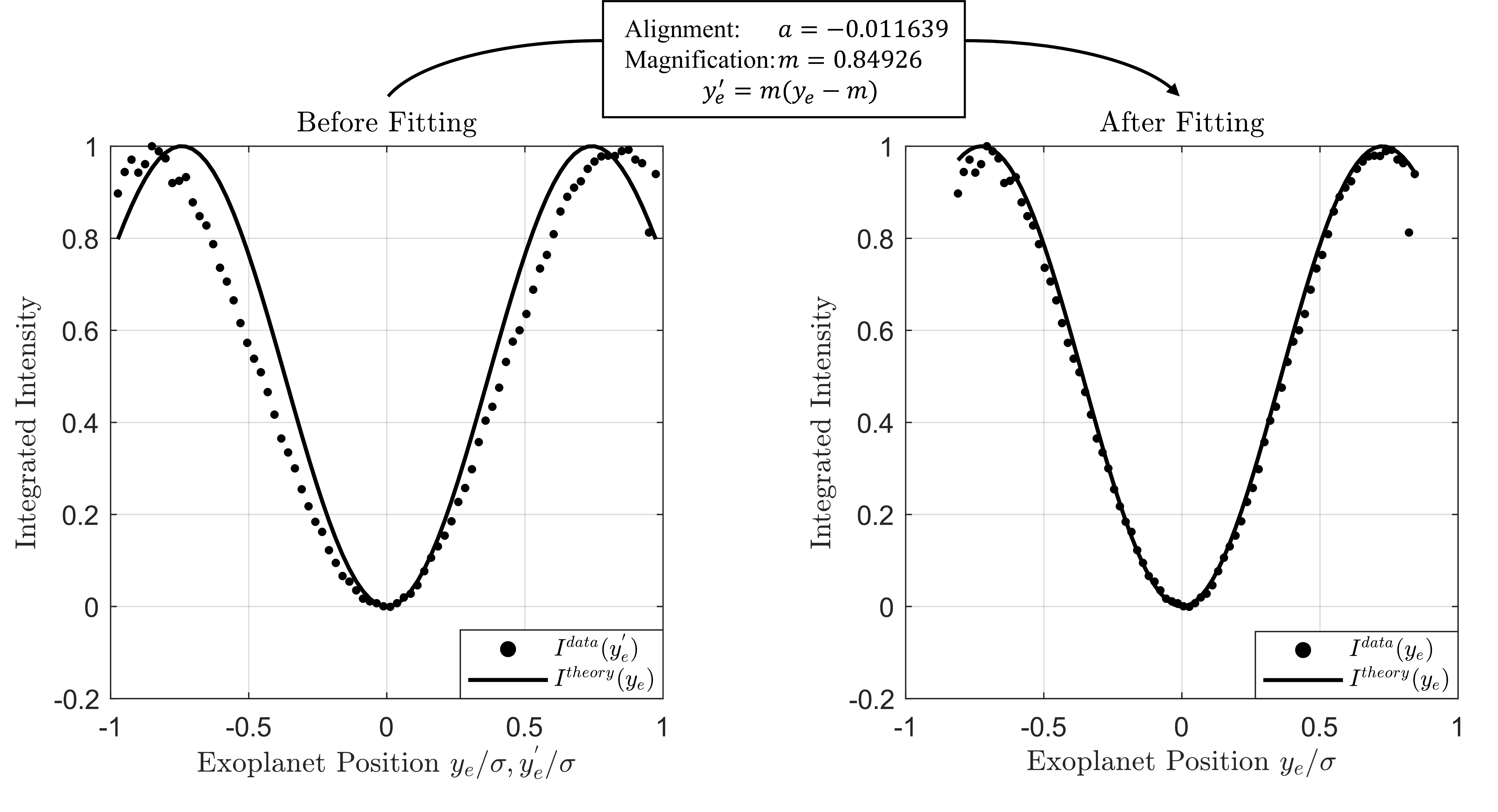}
    \caption{Alignment and magnification discrepancies in $4f$ system are accounted for by fitting the experimental throughput to theoretically-predicted values.}
    \label{fig: Bias-Scaling Fit}
\end{figure}

Figure \ref{fig: Bias-Scaling Fit} shows the experimental coronagraph throughput as a function of the scanned source position. Comparing the measured throughput to the theoretical throughput curve of our system revealed a scaling $m$ and shift $a$ between the true source displacements $y_e$ relative to the diffraction limit and the displacements computed from our system specifications $y_e^{'} = m(y_e - a)$. The scaling and shift terms arise from experimental error in magnification and alignment of the optical axis respectively. After accounting for these discrepancies, we see the throughput of curve of our transformed (scaled and shifted) coordinates align with theoretical predictions.

\section{Simulated MPLC Cross-Talk Matrix}
\label{apd: Simulated Cross-Talk}
In this section, we address the complex cross-talk matrix of the MPLC mode sorter in greater detail. Historically, literature in spatial mode sorting has placed emphasis on defining the cross-talk matrix in terms of the power leakage between modes (amplitude-only characterization). However, complete characterization of the cross-talk requires estimation of both the amplitude and phase. Direct experimental characterization of the complex cross-talk matrix is a critical to forthcoming high-performance mode-sorting applications such as coronagraphy and spatial mode filtering. In this section we study the cross-talk matrix of the MPLC used in our experimental design with numerical field propagation simulations.

The goal of an MPLC is to passively map a given set of input modes $\{{\psi}_i \in \mathbb{C}^M\}_{i=1}^{K}$ into a desired set of output modes $\{{\phi}_i\in \mathbb{C}^M\}_{i=1}^{K}$. For concreteness, let $\Psi,\Phi\in \mathbb{C}^{M\times K}$ be matrices whose columns are (vectorized) orthonormal input and output modes represented in the pixel basis. In general, the number of pixel degrees of freedom $M$ is much greater than the number of modes $K$ that one wishes to sort. Additionally, we take the modes to be orthonormal such that $\Psi^\dagger \Psi = \Phi^\dagger \Phi = I_{K}$. The cross-talk matrix $\Omega\in \mathbb{C}^{K\times K}$ describes the interferometric connection between $\Psi$ and $\Phi$ within the MPLC transfer matrix:
$$
T = \Phi\Omega^\dagger\Psi^\dagger
$$
Under this decomposition of the transfer matrix $T$, the action of the MPLC on an arbitrary incident field can be understood as follows: First, project the arriving field onto the basis of input modes via $\Psi^\dagger$. Then, interfere light in the input mode channels according to the cross-talk matrix $\Omega$. Finally, expand the mixed channels back into the pixel basis via $\Phi$. Note that if $\Omega=I_K$, then the MPLC perfectly maps the input modes to the output modes in a bijective fashion. In the most general case however, $\Omega$ admits a singular value decomposition which comprehensively handles interferometric mixing between mode channels as well as loss through the device.

\begin{figure}
    \centering
    \includegraphics[width=\linewidth]{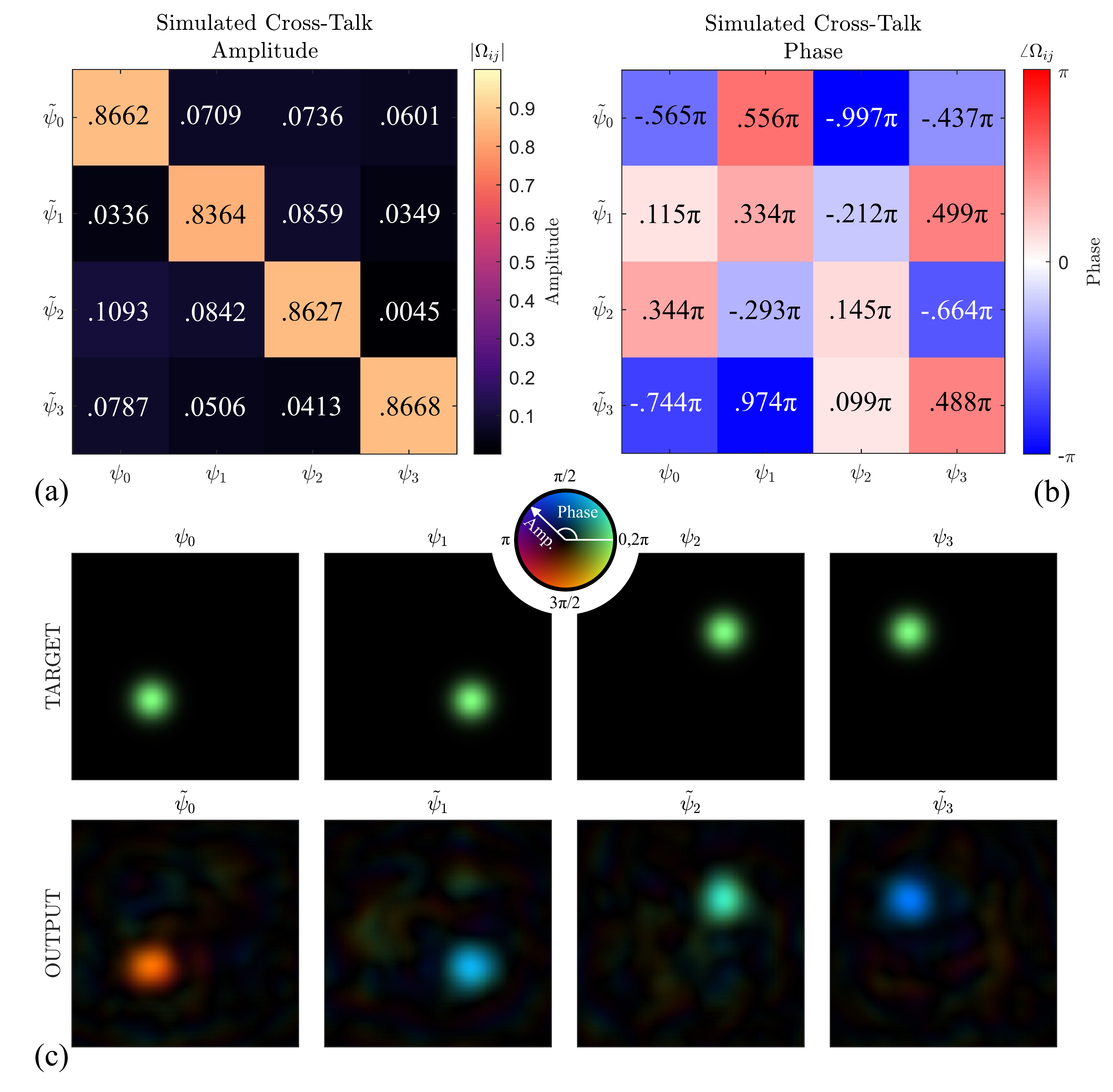}
    \caption{\textbf{(a)} Amplitude and \textbf{(b)} phase of the complex cross-talk matrix calculated from numerical simulations. \textbf{(c)} Simulated target and output fields of the mode sorter at the sorting plane after propagating the Zernike modes through the MPLC. The entries of the cross-talk matrix are the pair-wise projections (overlap integrals) between target modes and output modes.}
    \label{fig: Simulated Cross-Talk}
\end{figure}

A simple way to calculate the elements of the complex cross-talk matrix in simulation is by forward propagating each input mode independently and taking projection between the resulting output field at the sorting plane and the target output mode. That is we perform the computation,
$$
\Omega_{ij} = ({\phi}_i^\dagger T{\psi}_j)^{*} ,\qquad T = H_\tau D_\tau  \cdots H_1 D_1
$$
where we compute the transfer matrix via alternating application of phase masks (diagonal matrices $D_i$) inter-spaced by plane-to-plane light propagation (unitary matrices $H_i$). Figure \ref{fig: Simulated Cross-Talk}(a,b) shows the simulated cross-talk of our MPLC device. We note that the this simulated cross-talk matrix is \textit{not} unitary as the forward propagated modes span a slightly different subspace compared to the target mode set. This is visually evident in Figure \ref{fig: Simulated Cross-Talk}(c) where residual speckles in the output fields are observed to lie outside the designated regions of support for the target modes. Contrast this matrix to the experimental cross-talk matrix displayed in Figure \ref{fig: Experimental Cross-Talk} which we evaluated using the methods described in Appendix \ref{apd: Calibration}(1). Importantly, for calibrating the experimental cross-talk matrix, we assumed loss was negligible and forced the requirement that $\Omega$ be unitary. Despite this reduction in the degrees of freedom, the unitary cross-talk matrix calibrated for our bench-top implementation of the coronagraph provides an experimental model that aligns reasonably well with our measured results.

\section{Numerical Comparisons to Other Coronagraphs}
\label{apd: Coronagraph Comparisons}

\begin{figure}
    \centering
    \includegraphics[width = \linewidth]{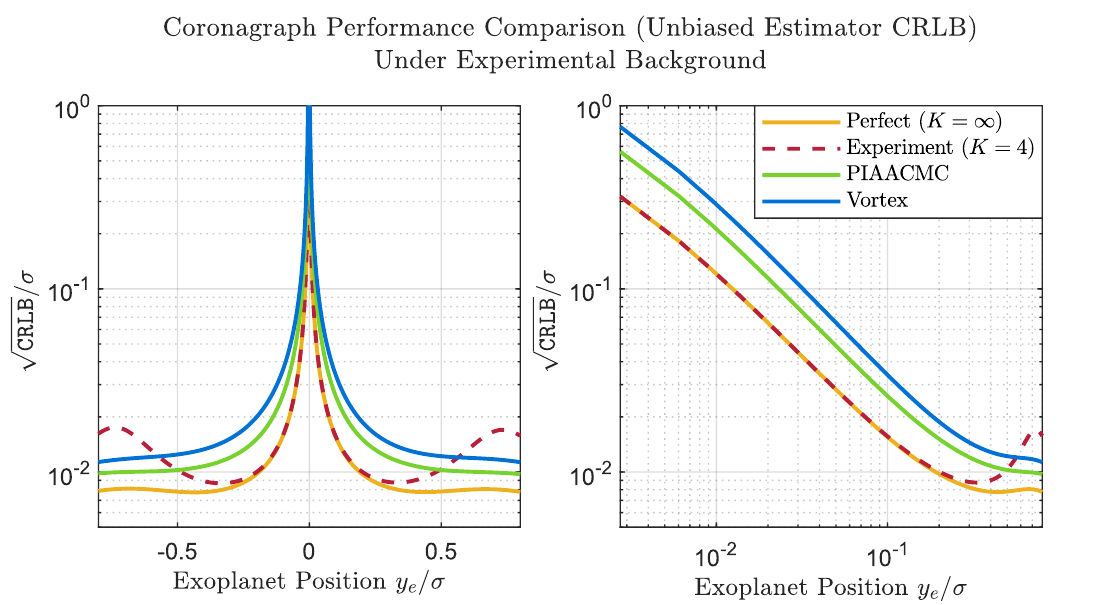}
    \caption{ Comparison of CRLBs for unbiased estimators of the exoplanet position under measurements with different coronagraphs. Each coronagraph measurement model includes the same empirical background noise characterized in our experimental system.}
    \label{fig: CRB Comparisons}
\end{figure}

As points of comparison, we numerically evaluate the performance of experimental coronagraph against the PIAACMC, Vortex (charge-2) coronagraph, and the Perfect coronagraph. The Perfect coronagraph is the idealized version of our experimental implementation in the absence of mode truncation or cross-talk. We assign the same background illumination profile espoused in our experiment $\Lambda_B \mbf{p}_B$ to the measurement model of each coronagraph. Then, the distinguishing factor between coronagraphs manifests in the term $\Lambda_0 \mbf{p}(\vec{r}_e)$ which now depends on singular-value decomposition of the field transformation performed by a given coronagraph,
\begin{equation}
    \mbf{q}(\vec{r}_0) = |U \Sigma V^{\dagger} \mbf{u}(\vec{r}_0)|^2
\end{equation}
where $U,V \in \mathbb{C}^{M \times M}$ are unitary matrices and $\Sigma \in \mathbb{R}^{M \times M}$ is a diagonal matrix of singular values. Figure \ref{fig: CRB Comparisons} shows the Cra\'mer-Rao lower bounds associated with each coronagraph. We see that the idealized version of our experimental setup (Experiment 4-Mode) outperforms both the PIAACMC and Vortex coronagraph for over sub-diffraction exoplanet locations in the range $|y_e| <= 0.4\sigma$ which corresponds to where the CFI of the truncated mode basis begins to depart from the QFI limit as shown in Figure \ref{fig: 4 Mode Zernike}(b).

\begin{figure}
    \centering
    \includegraphics[width=\linewidth]{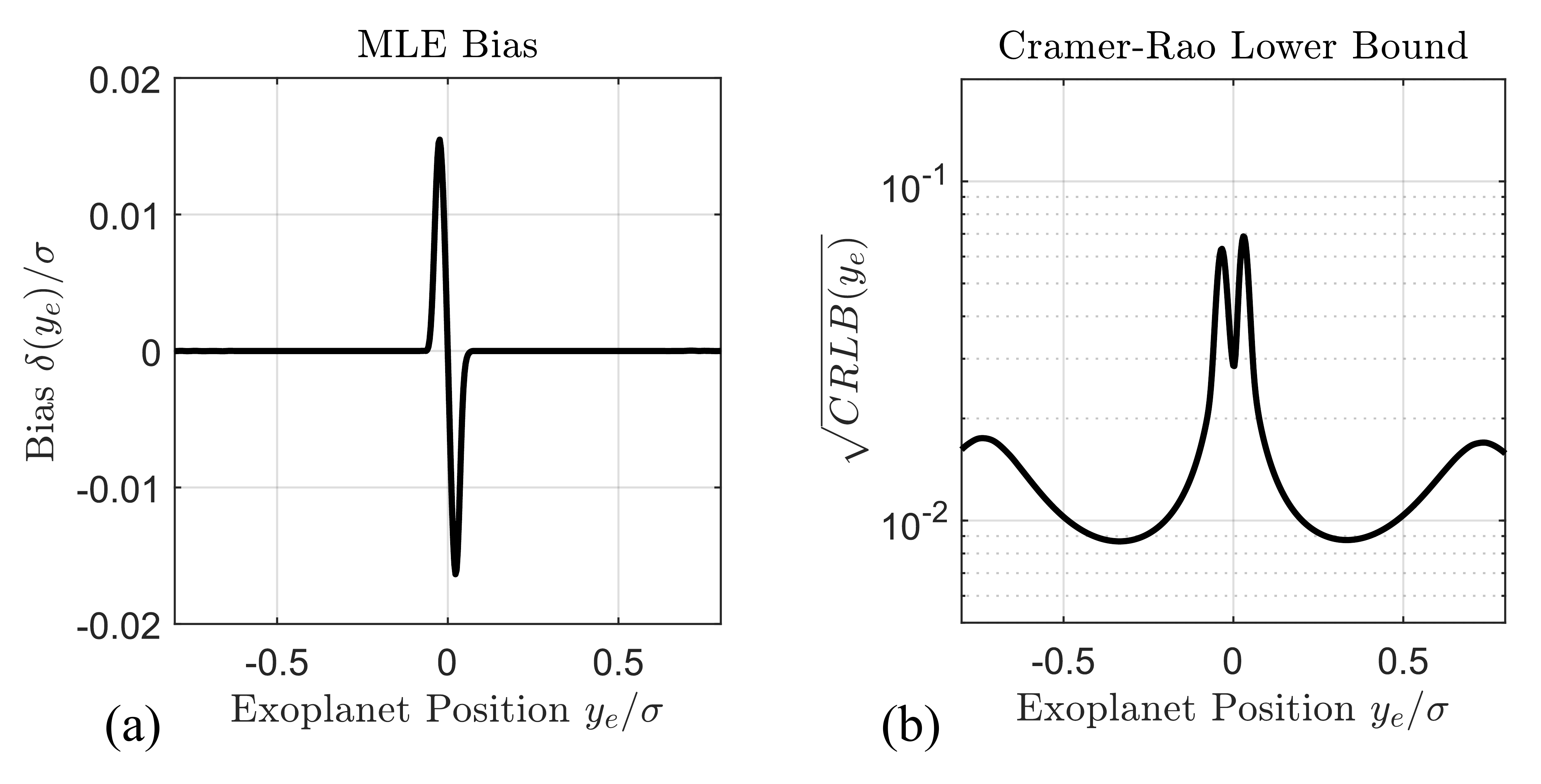}
    \caption{Numerical evaluation of the MLE bias and associated CRLB curve under experimental model. This CRLB curve is equivalent to that used in Figure \ref{fig: Error Analysis}(b)}
    \label{fig: MLE Bias}
\end{figure}

\section{Cramer-Rao Lower Bound for the Biased MLE}
\label{apd: CRLB Biased MLE}
The maximum likelihood estimator is known to be an unbiased estimator in the asymptotic limit of many repeated measurements (here the term 'measurement' formally refers to the detection of a photon). While this assumption is increasingly valid for larger star-planet separations, wherein the number of detected exoplanet photons is large, it becomes invalid for small star-planet separations as the vast majority of exoplanet photons are discarded by the coronagraph. This motivates the analysis of Equation \ref{eqn: CRLB} in the main text. The expectations are  $\mathbb{E}_{\mbf{Y}|
y_e}[\cdot]$ are taken over the joint distribution on the measurement outcome $\mbf{Y}$ given by,

\begin{subequations}
\begin{align}
    P(\mbf{Y}|y_e) &= \prod_{j=1}^{M}P(Y_j|y_e)\\
    P(Y_j = n|y_e) &= \frac{e^{-\tau_j(y_e)}\tau_j(y_e)^{n}}{n!}\\
    \tau_j(y_e) &= \Lambda_0 p_j(y_e) + \Lambda_B p_{B_j}
\end{align}
\end{subequations}
Finally, $I(y_e)$ is the total classical Fisher Information defined as,

\begin{equation}
    I(y_e) = \sum_{\mathcal{D}} P(\mbf{Y}|y_e) \bigg[ \frac{d}{dy_e}\log P(\mbf{Y}|y_e) \bigg]^2
\end{equation}
where $\mathcal{D}$ is the event space of possible measurement outcomes for $\mbf{Y}$. In Figure \ref{fig: MLE Bias}(a) we plot the bias of the MLE which we numerically computed by Monte-Carlo sampling the probabilistic measurement model of equation \ref{eqn: Measurement Model} to determine the expectation of the MLE $\mathbb{E}_{\mathbf{Y}|y_e}[\check{y}_{e}(y_e)]$. We see that near the optical axis, the MLE is clearly biased as it no longer operates in the asymptotic regime.  In Figure \ref{fig: MLE Bias}(b) we show the corresponding biased CRLB for the MLE of the exoplanet position.

\end{document}